\documentclass[times,twocolumn,final,authoryear]{elsarticle}

\usepackage{jasr}
\usepackage{framed,multirow}

\usepackage{amssymb}
\usepackage{latexsym}

\usepackage{algorithm}
\usepackage{algpseudocode}

\usepackage{url}
\usepackage{xcolor}
\definecolor{newcolor}{rgb}{.8,.349,.1}

\usepackage[citebordercolor=white]{hyperref}

\usepackage{subcaption}
\usepackage{textcomp, gensymb}
\usepackage{amsmath, amssymb}
\usepackage{footnote}
\usepackage{upgreek}
\usepackage{caption}

\usepackage{lineno}

\newcommand{\zpe}{22.15}

\newcommand{\noiseae}{-6.57}
\newcommand{\noisebe}{45.25}

\newcommand{\fwhm}{3.6}

\newcommand{\minvel}{0.043}
\newcommand{\maxvel}{0.380}

\journal{Advances in Space Research}

\begin{document}

\verso{Benjamin F. Cooke \textit{et al.}}

\begin{frontmatter}

\title{Simulated recovery of LEO objects using sCMOS blind stacking}%

\author[1,2]{Benjamin F. \snm{Cooke}\corref{cor1}}
\cortext[cor1]{Corresponding author: 
  benjamin.cooke@warwick.ac.uk;
  }
\author[1,2]{Paul \snm{Chote}}
\author[1,2]{Don \snm{Pollacco}}
\author[1,2]{Richard \snm{West}}

\author[1,2]{James A. \snm{Blake}}
\author[1,2]{James \snm{McCormac}}

\author[1,2]{Robert \snm{Airey}}
\author[1,2]{Billy \snm{Shrive}}

\address[1]{Department of Physics, University of Warwick, Gibbet Hill Road, Coventry CV4 7AL, UK}
\address[2]{Centre for Space Domain Awareness, University of Warwick, Gibbet Hill Road, Coventry CV4 7AL, UK}

\received{}
\finalform{}
\accepted{}
\availableonline{}
\communicated{}

\begin{abstract}
We present the methodology and results of a simulation to determine the recoverability of LEO objects using a blind stacking technique. The method utilises sCMOS and GPU technology to inject and recover LEO objects in real observed data. We explore the target recovery fraction and pipeline run-time as a function of three optimisation parameters; number of frames per data-set, exposure time, and binning factor. Results are presented as a function of magnitude and velocity. We find that target recovery using blind stacking is significantly more complete, and can reach fainter magnitudes, than using individual frames alone. We present results showing that, depending on the combination of optimisation parameters, recovery fraction is up to 90\% of detectable targets for magnitudes up to 13.5, and then falls off steadily up to a magnitude limit around 14.5. 
Run-time is shown to be a few multiples of the observing time for the best combinations of optimisation parameters, approaching real-time processing.
\end{abstract}


\end{frontmatter}


\section{Introduction}
\label{sec:Introduction}

\begin{figure}[htp]
    \centering
    \includegraphics[width=\columnwidth]{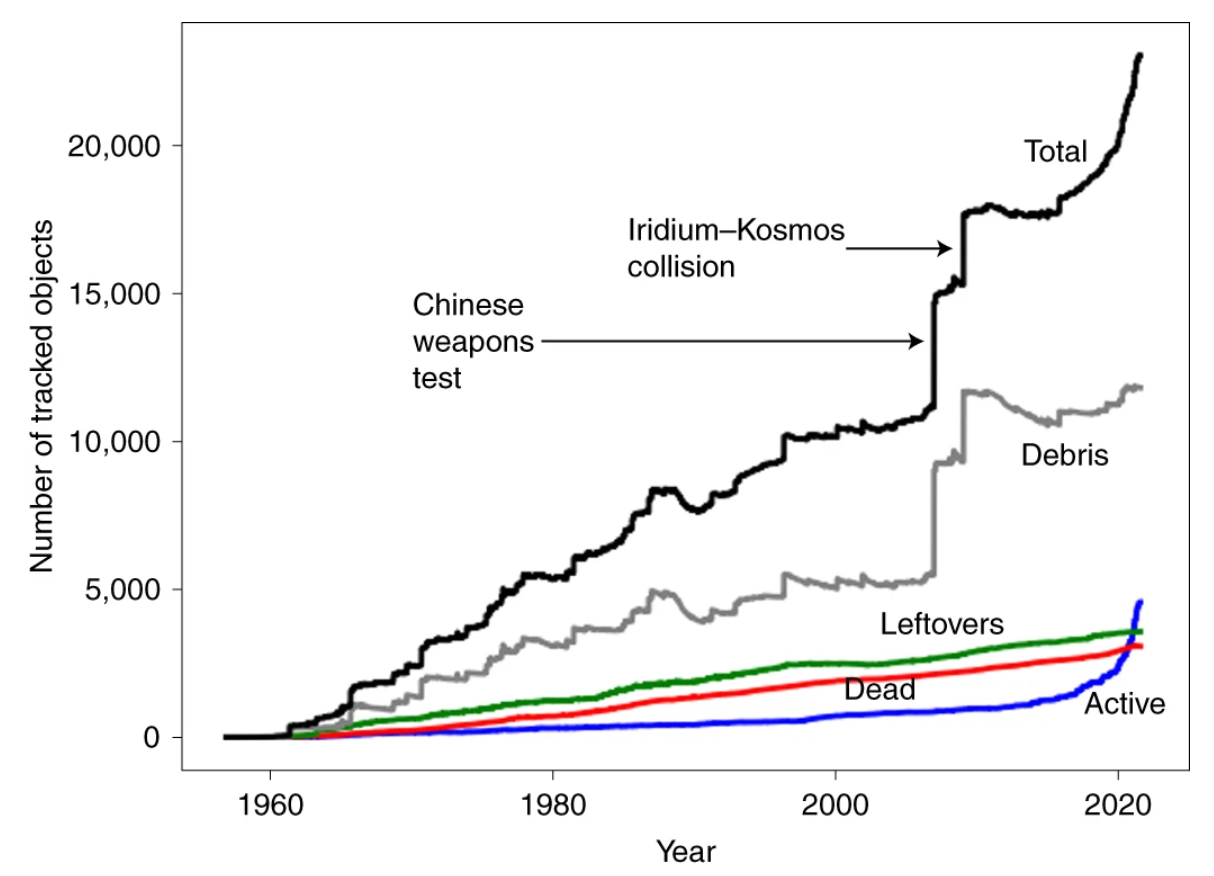}
    \caption{Number of catalogued objects in Earth orbit as a function of year. Reproduction of Fig. 1 from \citet{2022NatAs...6..428L}.}
    \label{fig:objects}
\end{figure}

In recent years the number of objects in Earth's orbit has increased dramatically \citep{blake2022looking,2021AcAau.184...11P}. This is due to both launches, including a number of large satellite constellations (e.g. Starlink, OneWeb, etc.) \citep{BERNHARD2022}, as well as the result of unintentional collisions (e.g. the Iridium 33 and Cosmos 2251 collision in 2009 which produced approximately 2300 catalogued fragments) and intentional anti-satellites tests (e.g. the Chinese missile test in 2007 which produced more than 2000 trackable pieces of debris) \citep{2010amos.confE..37S}. Fig. \ref{fig:objects} shows how this number has changed since the start of the space exploration era.

The number of fully characterised and catalogued Low Earth Orbit (LEO) objects is approximately 80\% for objects $\geq10$cm but much less for objects smaller than this \citep{2021AcAau.184...11P}. With an average collision velocity of 10km/s, even a 1cm sized LEO object can cause a fragmentation event should it collide with another object \citep{2020AdSpR..65..351O}. Every collision has the potential to lead to not only the loss of satellite functionality, but also the chance for fragments to produce a cascading collision cycle leading to an eventual Kessler syndrome in which part of LEO becomes unusable \citep{1978JGR....83.2637K}. These factors mean that the need for more robust and adaptable tracking of LEO satellites and debris is vital.

Most optical detection methods rely on either identification from single images \citep{chote2019precision,2018AcAau.145..332D}, or must make assumptions about the orbit of an object before observation \citep{blake2021debriswatch}. In this way a telescope can track the assumed orbit, reducing the relative motion of the target and thus increasing its detectability. The obvious downsides of this method are that every different orbit requires a different data-set and only targets on, or close to, the observed orbit are detectable. The method laid out in this paper, blind stacking, is designed to mitigate these effects \citep{10.1093/pasj/psab017,2020PSJ.....1...81R}. Blind stacking, also known as track-before-detect, 
does not track any specific orbit therefore we have freedom to choose the preferred tracking rate. For simplicity of design this work employs a fixed orientation instrument. Blind stacking can be used to locate targets on any given orbit, moving in any chosen direction and can identify multiple targets on different orbits within the same data-set, provided the object spends sufficient time within the instrument field of view.

Of course, blind stacking also has its deficits. When compared to traditional stacking, blind stacking has a reduced sensitivity; since the light from a particular target 
is spread over more pixels, the corresponding surface brightness is reduced. Blind stacking is also a multi-variable problem with exposure time, plate scale, observation angle, velocity range, number of paths and number of frames all having an impact on the recoverability of LEO targets. This paper attempts to quantify some of these parameters and explores blind stacking as an optical LEO object observational tool.

To reach the precisions and exposure times required for the optical detection of objects at LEO this simulation uses scientific Complementary Metal–Oxide–Semiconductor (sCMOS) detectors. 
Significant advances in sCMOS detectors in recent years have made them competitive with traditional CCD detectors, and in applications where high-framerates are required they are clearly superior \citep{2020AAS...23517501W}. These detectors can have extremely high quantum efficiencies, upwards of 90\%, as well as very fast read-out times while maintaining a very low read-out noise (typically on the order of 1-2 electrons). This means sCMOS detectors allow us to observe continuously with sub-second exposure times without sacrificing precision \citep{10.1093/pasj/psac035,2016amos.confE..25Z}. These aspects make sCMOS detectors ideal for LEO observations where the targets are moving quickly and thus require short exposure times, or large fields of view, to capture and characterise \citep{chote2019precision,2019LPICo2109.6207W}. The high cadence of observations enabled by the use of sCMOS devices, combined with the high computing resources required by the blind stacking method, make this problem unfeasible for traditional CPU-based computing. Fortunately however, the problem is highly parallelisable, meaning the computational time can be drastically reduced through judicious use of GPU processing. GPUs are well suited for this purpose and can result in multiple orders of magnitude improvements in computational runtime for appropriate algorithms.

This paper proceeds as follows. Section \ref{sec:Methodology} lays out the data acquisition and the set-up and execution of the simulation. Section \ref{sec:Results} details the key results and Section \ref{sec:Discussion and conclusions} analyses said results and presents our conclusions.

\section{Methodology}
\label{sec:Methodology}

\subsection{Data acquisition and reduction}
\label{sec:Data acquisition and reduction}

\begin{figure}[htp]
    \centering
    \includegraphics[width=\columnwidth]{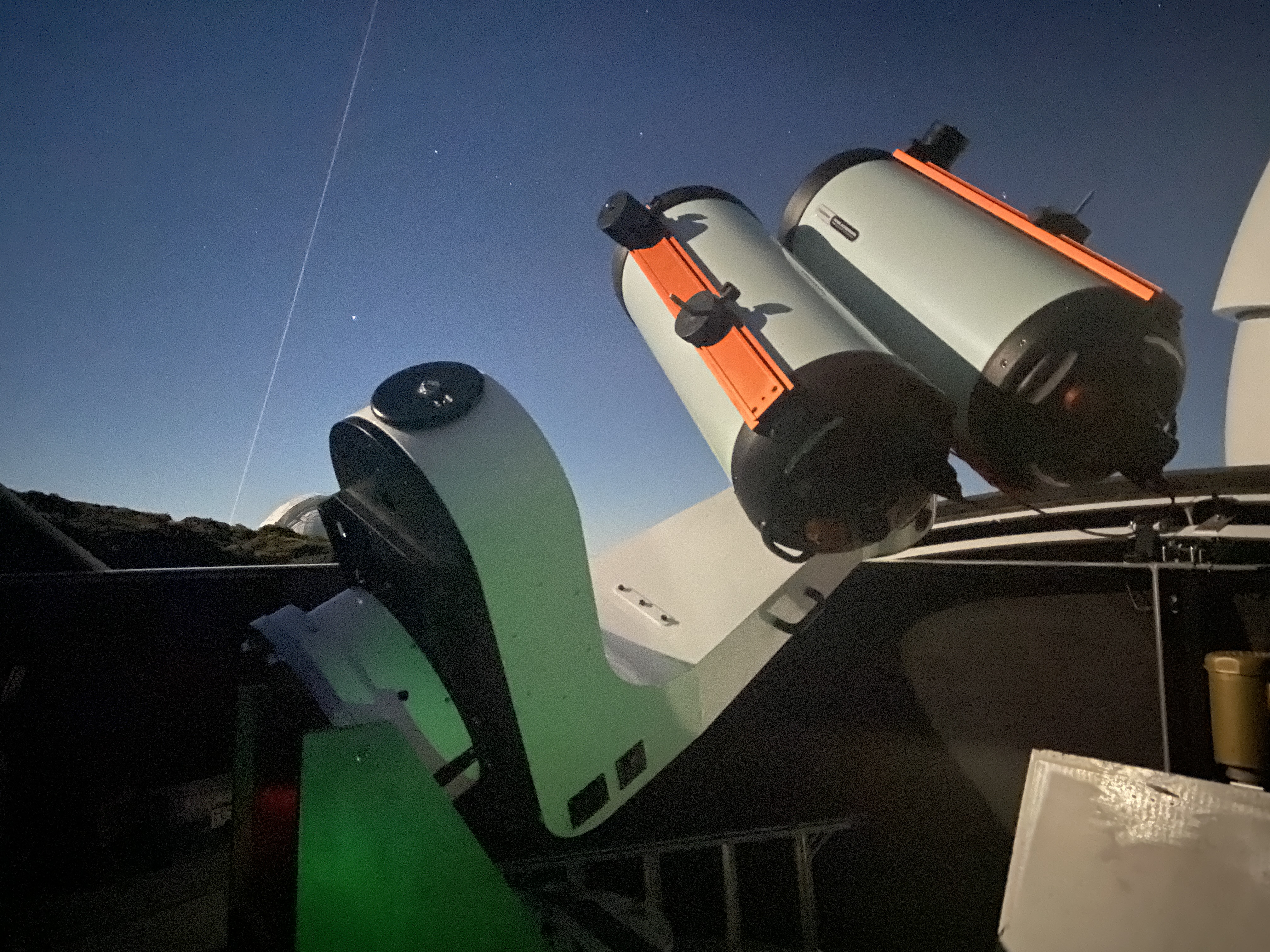}
    \caption{Warwick CLASP telescope on Roque de los Muchachos Observatory, La Palma. Photo credit: Paul Chote.}
    \label{fig:CLASP}
\end{figure}

The data used in these simulations were obtained using the University of Warwick CLASP test telescope, located at the Roque de los Muchachos Observatory on La Palma in the Canary Islands. The telescope features two 36cm F/2.2 prime focus RASA 36 astrographs mounted on a direct drive Planewave L600 mount with 61MPix 3.76 micron pixel QHY600M sCMOS detectors. Each instrument has a field of view of $2.63 \times 1.76$\,deg. The zero point of the system, i.e. the magnitude of a target which produces a flux of 1 e$^-$/s, is found to be \zpe\,mag. The typical Full Width at Half Maximum (FWHM) of the system Point Spread Function (PSF) is determined using a combination of Source Extractor Python (\texttt{SEP}) and \texttt{SciPy} \citep{2016JOSS....1...58B,1996A&AS..117..393B,2020SciPy-NMeth} and found to be $\sim$\fwhm\,pixels ($\sim$0.001\,deg). 
Table \ref{tab:CLASP} summarises the instrument specifications and Fig. \ref{fig:CLASP} shows the instrument set-up.

\begin{table}[htp]
\centering
\begin{tabular}{c|c}
Parameter    & Value                           \\ \hline
Diameter     & 36\,cm                          \\
FoV          & $2.63 \times 1.76$\,deg \\
Focal length & F/2.2                           \\
Detector     & QHY600M sCMOS                   \\
Pixel size   & 3.76\,$\upmu$m   \\
Gain         & 0.42\,e$^-$/ADU                 \\
FWHM         & 3.6\,pixels                     \\
Zero point   & \zpe\,mag       
\end{tabular}
\caption{CLASP instrument specifications.}
\label{tab:CLASP}
\end{table}

Data were obtained over a range of nights and at a range of exposure times from 0.125\,sec to 1.0\,sec. Pixel values are currently read out over a fibre connection which limits the maximum cadence to 0.253s\,sec. Shorter exposures are possible but currently result in deadtime where photons are not collected. This limit is temporary and greater cadence values are included in the following simulation to account for planned improvements in the system. Significant improvements in cadence however, would require smaller image sizes, or an adapted observational set-up, avenues which are being considered. Image frames are reduced using master bias and master flat fields and background subtracted using 
SEP. Pixel values are capped at the $99^{\rm th}$ quantile to reduce the impact of particularly hot or noisy pixels, as well as to limit the impact of bright background stars.

Observations are taken with a stationary telescope, resulting in trailing objects as well as trailing stars. Observing with a sidereally tracking system was considered, since it would reduce the length of star contaminated pixels in individual images. However, we find that, for the telescope system used, and the relevant binning and exposure times employed, the reduction in star trail length in pixels is fairly negligible. Therefore, it was decided to proceed with a completely stationary system to limit the number of moving parts and ensure the results are applicable to the simplest observing strategy possible.

\begin{figure}[htp]
    \centering
         \includegraphics[width=\columnwidth]{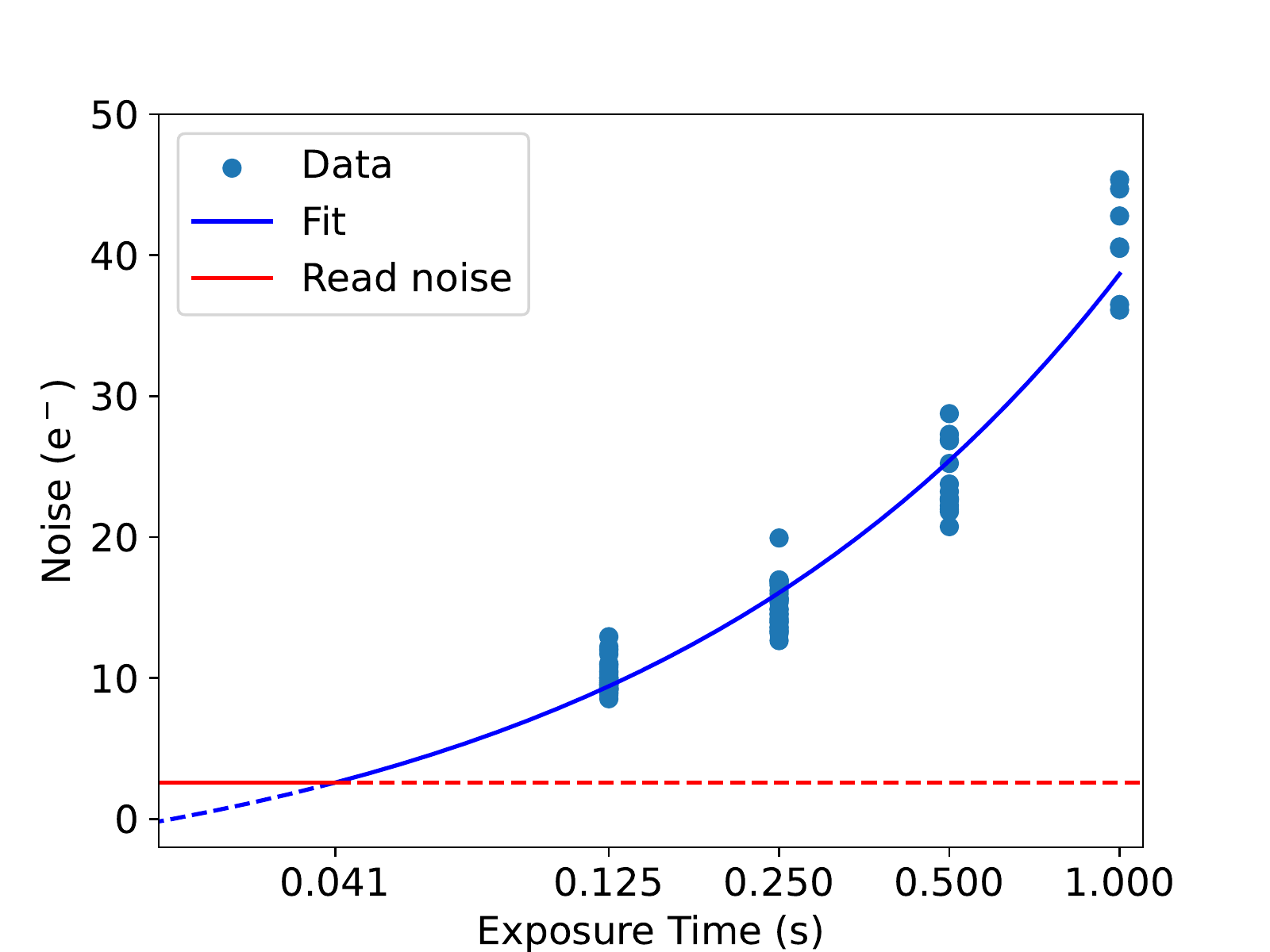}
    \caption{
    Plot showing how the noise properties of individual images change with exposure time. Also included is a line of best fit and the read noise limiting value. The transition to being read noise limited occurs at an exposure time of 0.041s.}
    \label{fig:zp_noise}
\end{figure}

From these data we can calculate 
the background noise as a function of exposure time. Fig. \ref{fig:zp_noise} shows these details.

From this plot we see that 
the background noise increases as the square root of the exposure time. This is due to the fact that longer exposures increase the number of background sky photons and therefore lead to an increase in photon shot noise. This aspect leads to a favouring of short exposure times when attempting to probe fainter targets. At the shortest exposures the noise eventually hits the read noise limit of 2.6 electrons (however, this happens below any of the exposure times considered in this paper). 
The 
background noise is given by

\begin{equation}
        \rm noise = 
\begin{cases}
    \noiseae + \noisebe\sqrt{t} & \text{if } t\geq 0.041\\
    2.6              & \text{otherwise.}
\end{cases}
\end{equation}


\subsection{Target velocities}
\label{sec:Target velocities}

\begin{figure}[htp]
    \centering
    \includegraphics[width=\columnwidth]{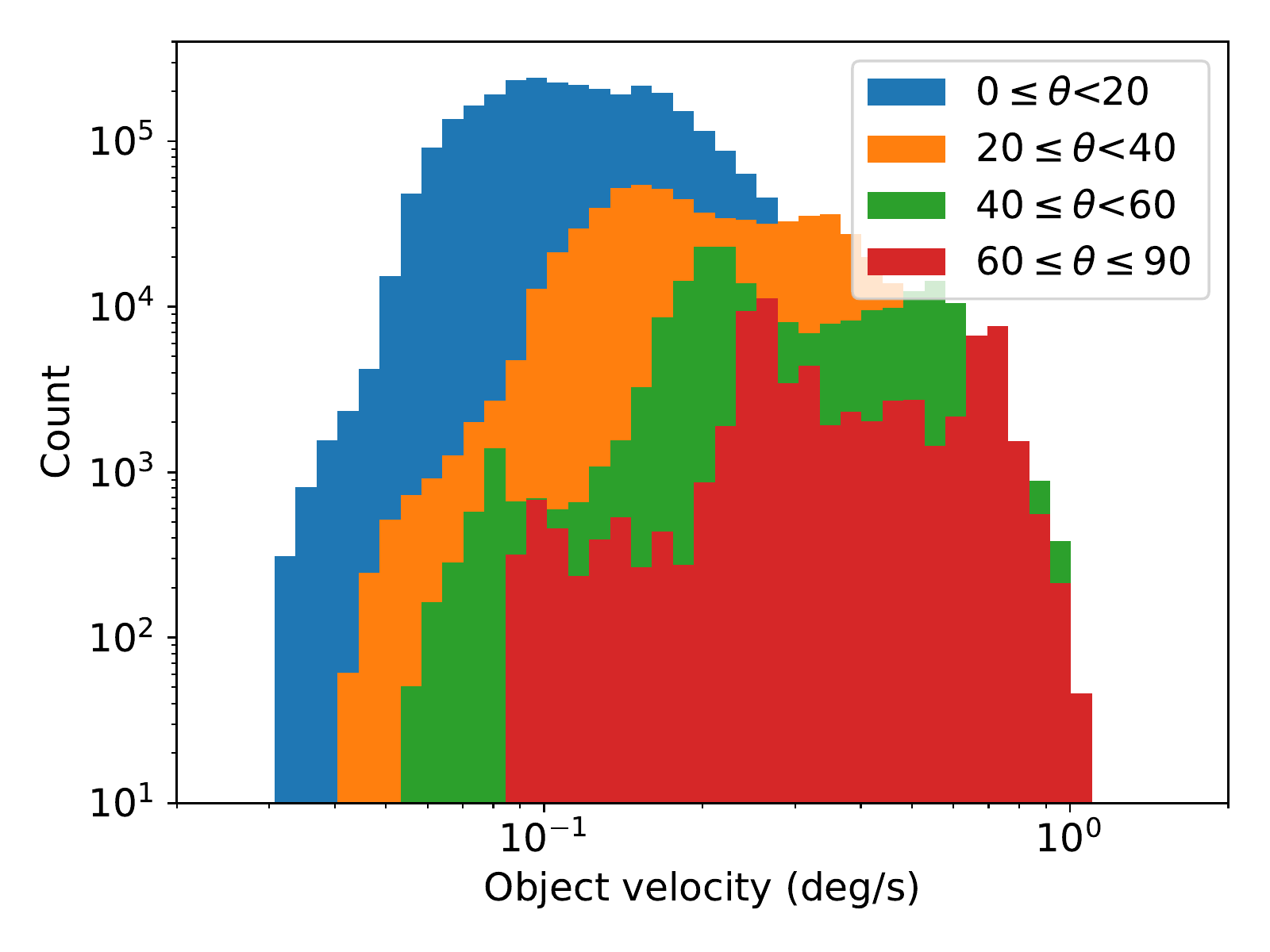}
    \caption{Velocity of satellites as observed from La Palma on a chosen night, separated by observational elevation angle, $\theta$, in degrees. Only sunlit satellites are included.}
    \label{fig:velocities}
\end{figure}

To best utilise the blind stacking procedure the optimal velocity range of targets must be known. That is, it is beneficial to search only the range of velocities that a target might reasonably have, since testing more paths increases both the runtime and the background noise of a stacked image (see Section \ref{sec:Blind stacking}). To determine this range we choose a random date and download all recent LEO Two-Line Elements (TLEs) from the Space Track database (space-track.org). LEO objects are selected using mean motion and eccentricity limits. Recent in this case means having been updated within the last 20 days. For each object in this catalogue we calculate its location and velocity, as observed from La Palma, at regular intervals throughout an entire night. This distribution of velocities gives a reasonable range of velocities to target using the blind stack search. Fig. \ref{fig:velocities} shows the distribution of velocities in deg/sec for all of the catalogue targets. 
Since the velocity of a target is lower when it is closer to the observer's horizon, we separate the plot into four observational elevation angle bands (denoted $\theta$, given in degrees). Additionally, only sunlit targets are included since these are the only ones visible to our optical system. Repeating this search for different dates resulted in negligible variances.

For the velocity range taken for this simulation we focus only on targets which are both sunlit and at least 20\,deg above the horizon, meaning they are observable with our current telescope setup. Since this distribution has a high-velocity tail (note the log-log scale) our upper limit is taken to be the $90^{\rm th}$ quantile of the valid targets. This means that we will not be able to recover the fastest 10\% of targets but, due to the number of paths scaling with velocity squared (see Section \ref{sec:Blind stacking}) we save approximately 90\% of the computational time associated with testing paths (the reduction in tested paths also results in a reduced background in the stacked images). The selected range is \minvel-\maxvel\,deg/sec. Limiting observations to lower observational elevation angles only could reduce this range further. Observing at lower elevation angles results in additional challenges, one of the most important being an increase in the average distance between the target and the observer. The increased distance leads to a reduction in target brightness and thus a further difficulty in target recovery. Here, we consider recoverability as a function of magnitude of the target as seen by the observer, not an absolute value, so our recoverability limits are independent of observational elevation angle. However, it should be noted that the same magnitude target will correspond to a larger physical target size when observed at lower elevation angles. Additionally, factors such as increased sky-mass and reduced seeing occur at lower observation elevation angles. These will affect recoverability but, since this simulation does not separate targets by elevation angle (other than to remove the very lowest targets) these impacts are beyond the scope of this work.

\subsection{Simulated signals}
\label{sec:Simulated signals}

We next had to simulate a range of realistic target signals that could be tested for recovery using our blind stacking code. Based on the acquired data discussed in Section \ref{sec:Data acquisition and reduction} we determined that the PSF of the system had a FWHM of \fwhm\,pixels. To simulate the object streak we first chose a velocity from the distribution above. This velocity, combined with the exposure time, give the length of the streak in a single image. The streak was simulated on a grid of pixels with dimensions equal to the streak length plus a buffer zone of 3 times the FWHM. The streak was split into multiple sections and at each step the location of the target was calculated, assuming that it moved uniformly from the start of the streak to the end over the duration of the exposure time. The total motion of an object across the field of view is only a few degrees so curvature is negligible. At each step in time the distance from each pixel to the target location was calculated and used to determine its flux based on a normal distribution with 
$\sigma=\rm FWHM/(2\sqrt{2\ln{2}})$. This was repeated for all time steps with the flux at each pixel being summed, resulting in a diffuse streak. The resulting image was then scaled, according to the zero point and exposure time to give the flux of a target of any specific magnitude using the following equation:

\begin{equation}
    F = 10^{(Z-M)/(2.5)}t,
\end{equation}

where $F$ is the total electron flux, $Z$ is the zero point of the data, $M$ is the required magnitude and $t$ is the exposure time. To convert flux from e$^-$ to ADU we divide by the nominal gain of 0.42e$^-$/ADU.

\begin{figure}[htp]
    \centering
    \begin{subfigure}{\columnwidth}
        \centering
        \includegraphics[width=\textwidth]{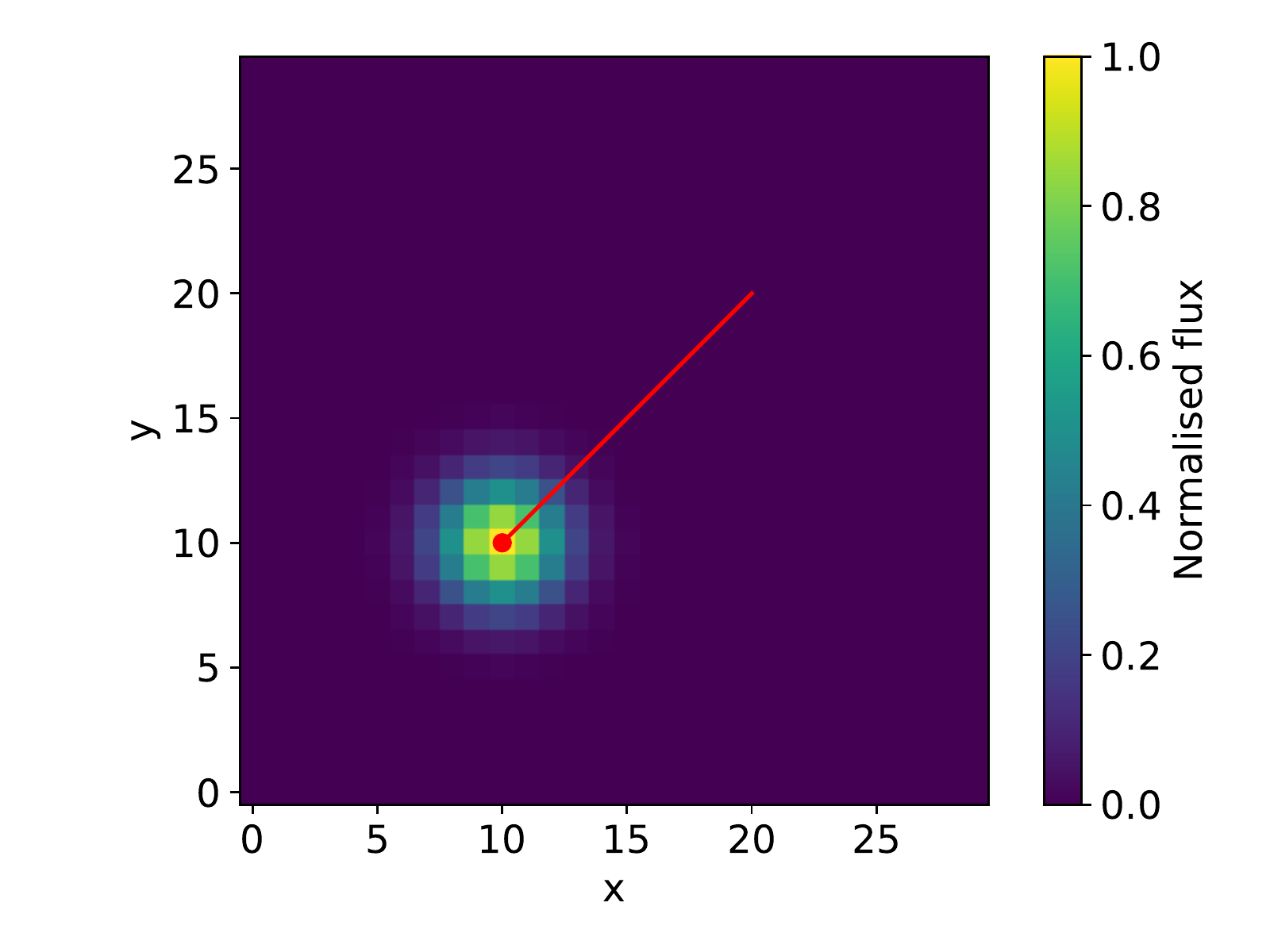}
        \caption{Simulated streak comprised of one point distribution.}
        \label{fig:streak_generation_sequence1}
    \end{subfigure}
    \begin{subfigure}{\columnwidth}
        \centering
        \includegraphics[width=\textwidth]{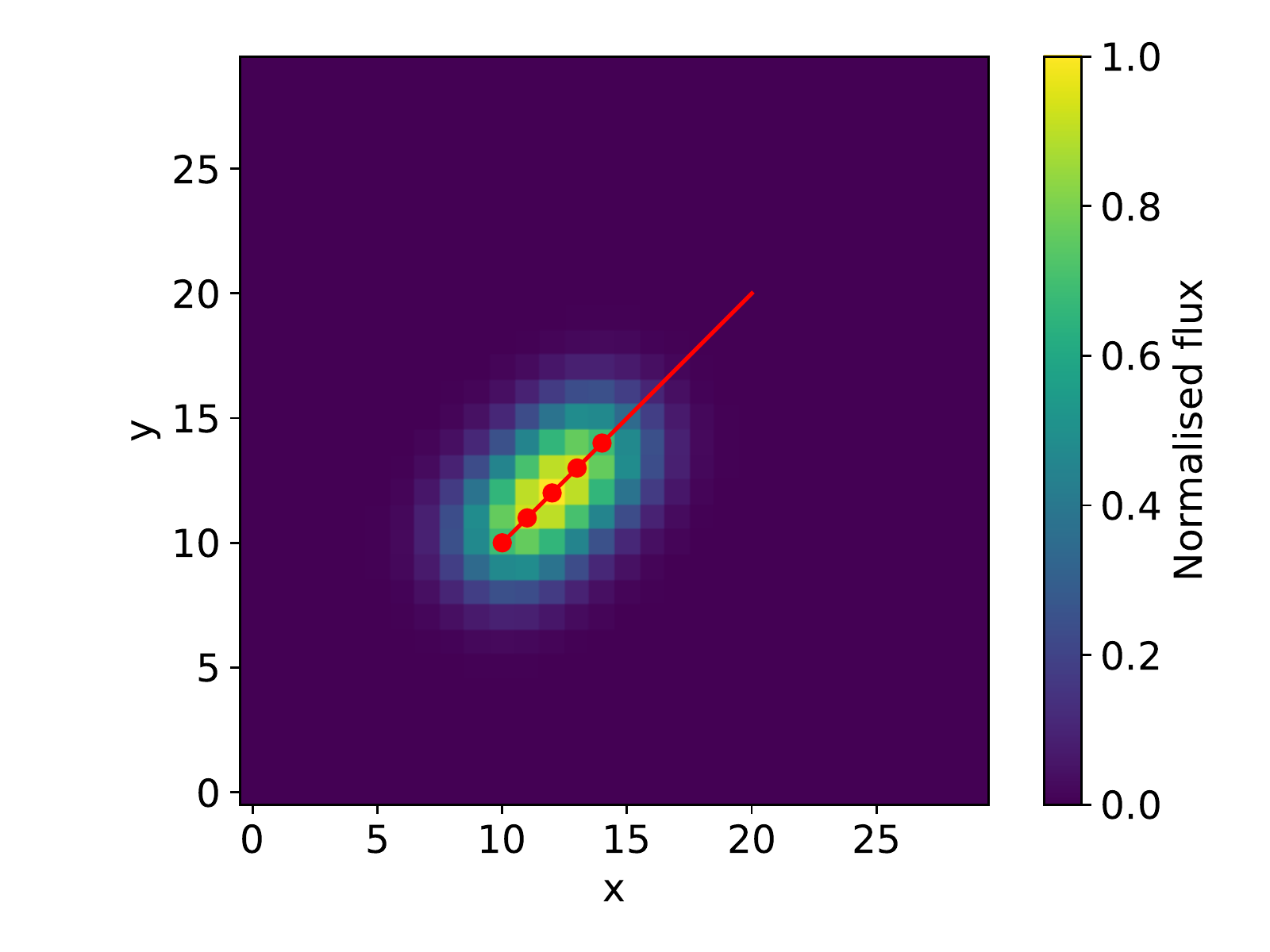}
        \caption{Simulated streak comprised of five point distributions.}
        \label{fig:streak_generation_sequence2}
    \end{subfigure}
    \begin{subfigure}{\columnwidth}
        \centering
        \includegraphics[width=\textwidth]{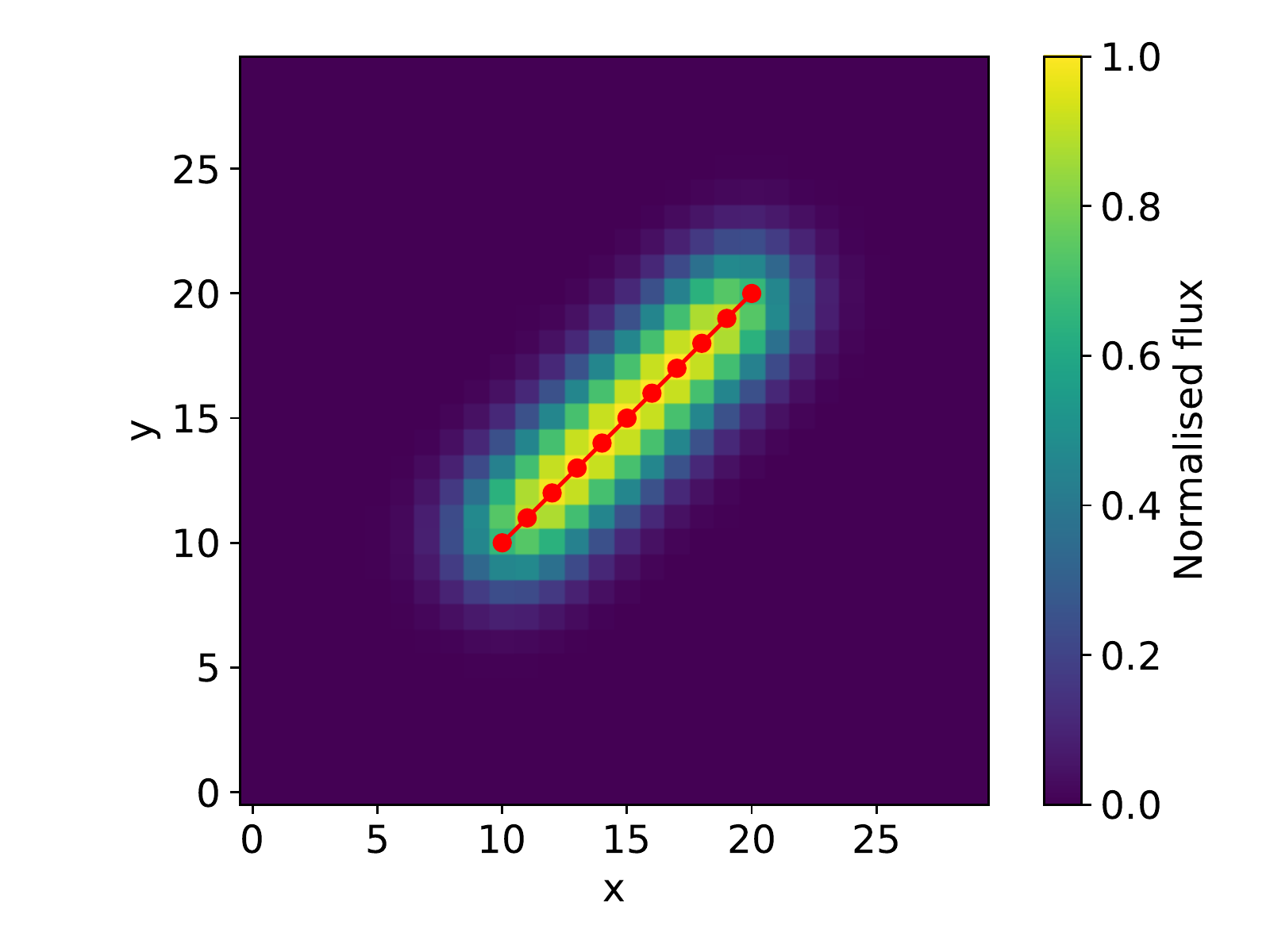}
        \caption{Simulated streak comprised of eleven point distributions.}
        \label{fig:streak_generation_sequence3}
    \end{subfigure}
    \caption{Generation of a simulated target streak, built up from multiple point distributions. The red line shows the targets motion and the red points shows the centres of the individual point distributions.}
    \label{fig:streak_generation_sequence}
\end{figure}

Fig. \ref{fig:streak_generation_sequence} shows the generation of an example target streak from a combination of multiple point distributions. The red line shows the motion of the target across a single frame with the red points showing the centres of the individual point distributions that are combined to create the streak effect.

This streak could then be generated for a random angle and injected into any pixels in the individual data frames. The position in the initial frame was chosen at random with its location in subsequent frames determined by the objects velocity, direction and frame exposure time. 
Objects that moved off the edges of the image between frames were no longer injected, so as to simulate targets which are not observed for the full data-set.

\begin{figure}[htp]
    \centering
    \begin{subfigure}{0.3\columnwidth}
        \includegraphics[width=\textwidth]{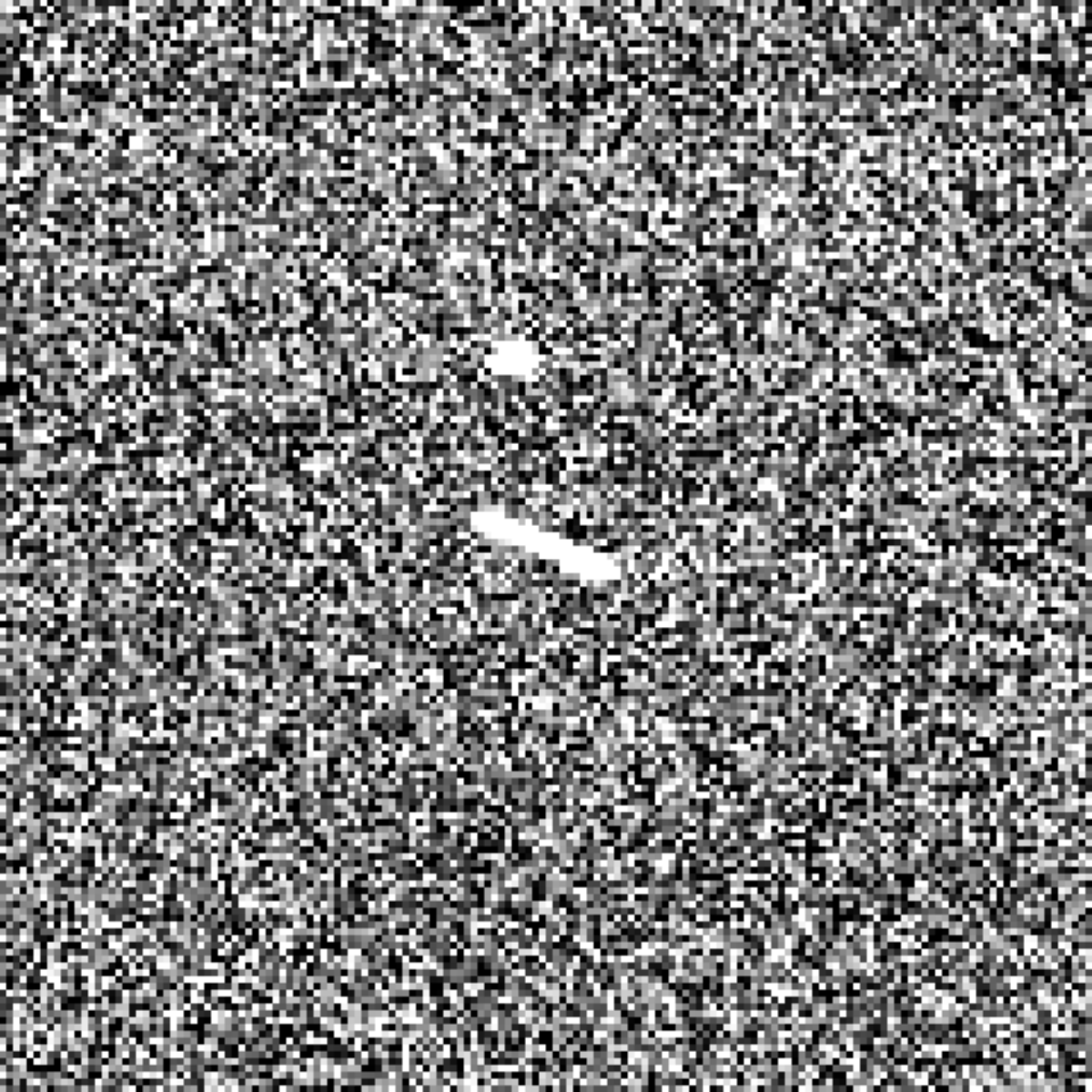}
        \captionsetup{format=hang}
        \caption{0.125\,s exposure\newline binning factor 2\newline 11$^{\rm th}$ mag target}
        \label{fig:example_frame_a}
    \end{subfigure}
    \begin{subfigure}{0.3\columnwidth}
        \includegraphics[width=\textwidth]{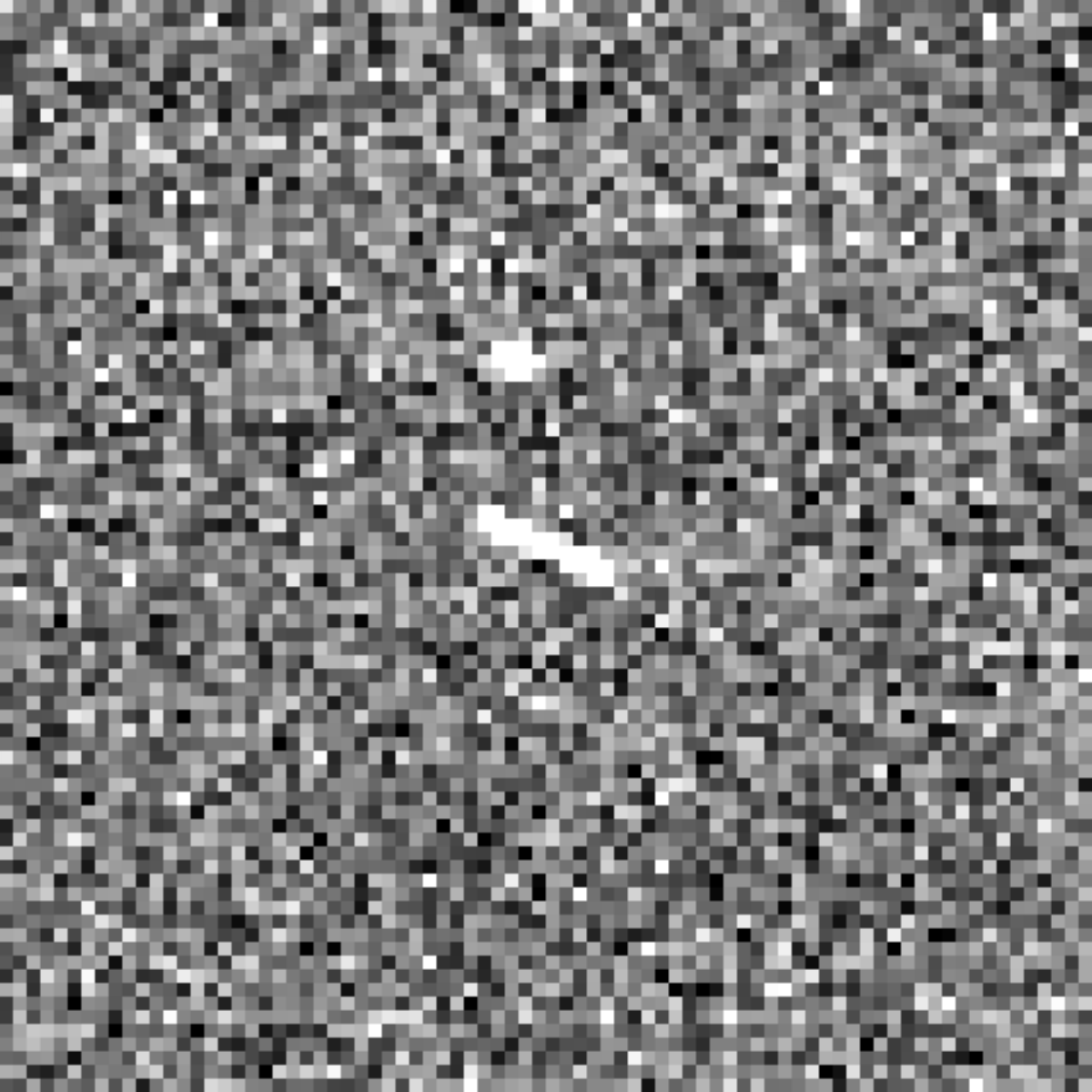}
        \captionsetup{format=hang}
        \caption{0.125\,s exposure\newline binning factor 4\newline 11$^{\rm th}$ mag target}
        \label{fig:example_frame_b}
    \end{subfigure}
    \begin{subfigure}{0.3\columnwidth}
        \includegraphics[width=\textwidth]{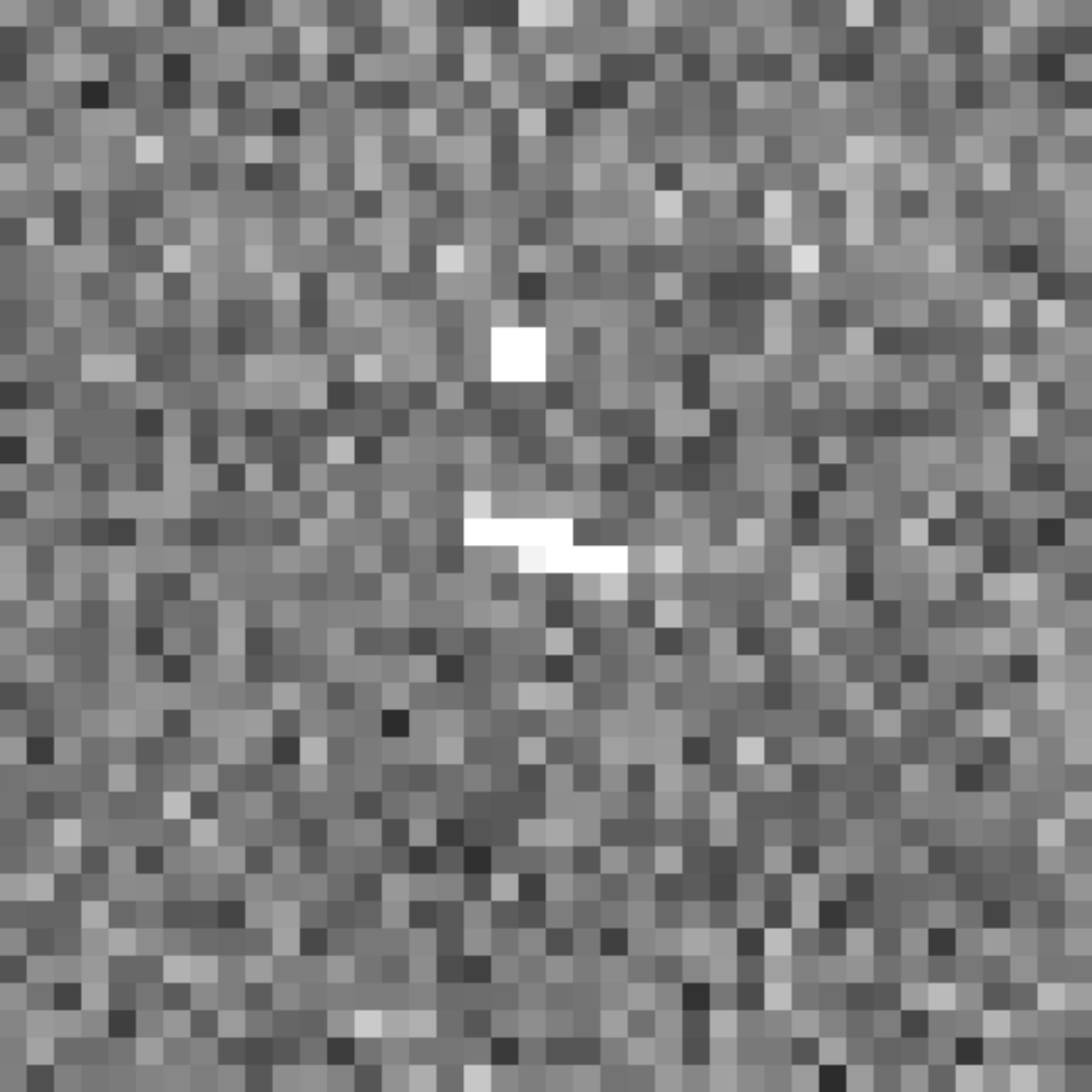}
        \captionsetup{format=hang}
        \caption{0.125\,s exposure\newline binning factor 8\newline 11$^{\rm th}$ mag target}
        \label{fig:example_frame_c}
    \end{subfigure}
    \begin{subfigure}{0.3\columnwidth}
        \includegraphics[width=\textwidth]{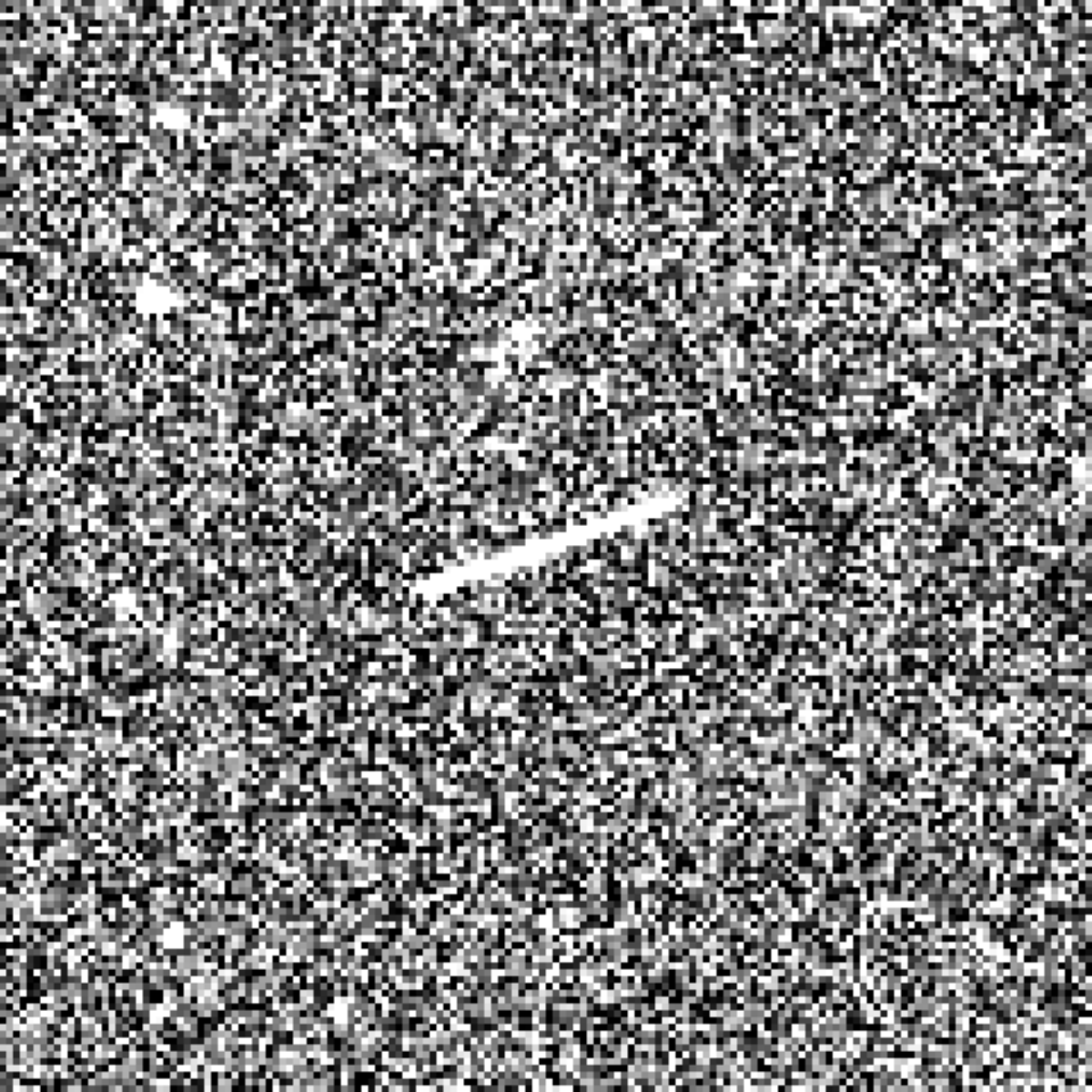}
        \captionsetup{format=hang}
        \caption{0.250\,s exposure\newline binning factor 2\newline 12$^{\rm th}$ mag target}
        \label{fig:example_frame_d}
    \end{subfigure}
    \begin{subfigure}{0.3\columnwidth}
        \includegraphics[width=\textwidth]{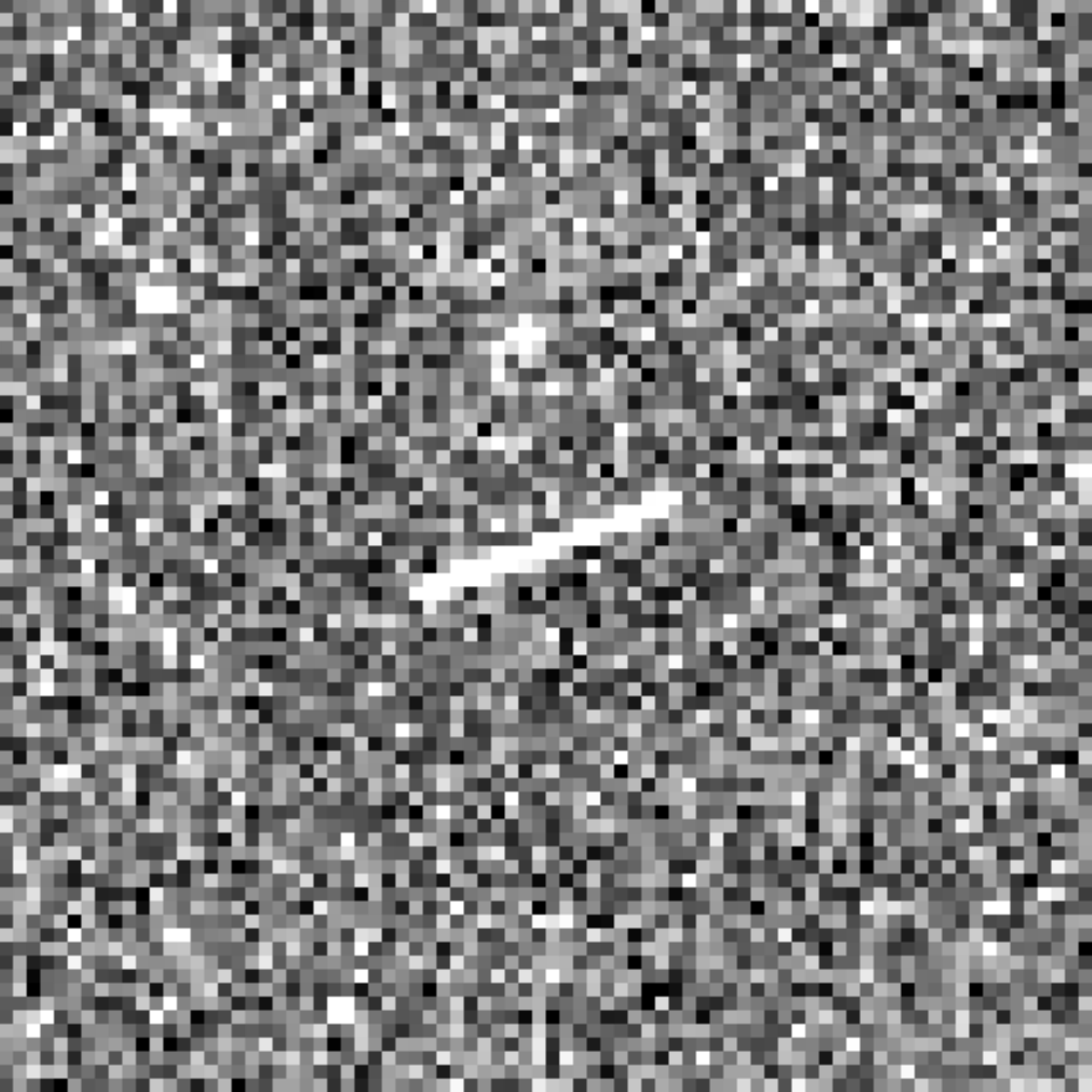}
        \captionsetup{format=hang}
        \caption{0.250\,s exposure\newline binning factor 4\newline 12$^{\rm th}$ mag target}
        \label{fig:example_frame_e}
    \end{subfigure}
    \begin{subfigure}{0.3\columnwidth}
        \includegraphics[width=\textwidth]{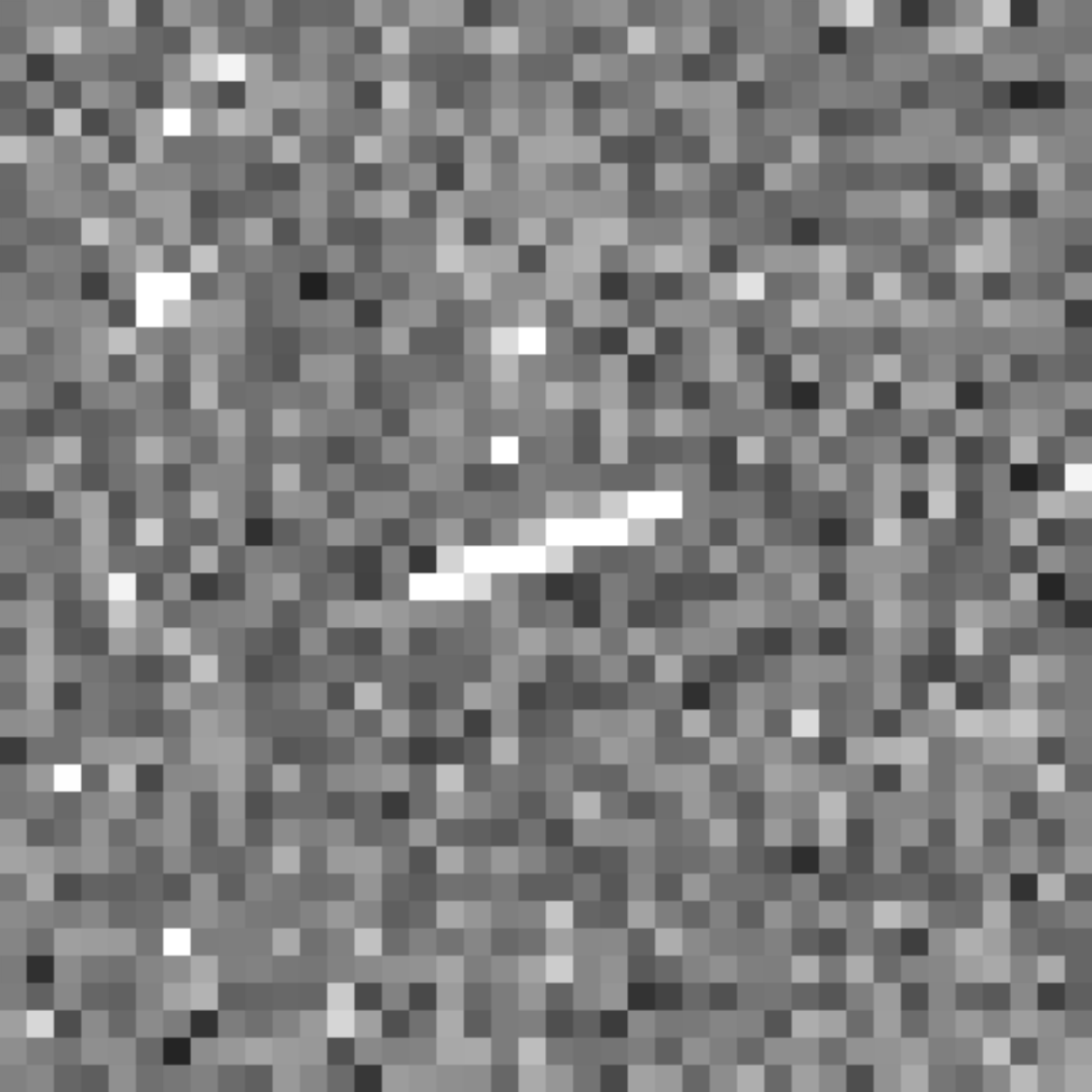}
        \captionsetup{format=hang}
        \caption{0.250\,s exposure\newline binning factor 8\newline 12$^{\rm th}$ mag target}
        \label{fig:example_frame_f}
    \end{subfigure}
    \begin{subfigure}{0.3\columnwidth}
        \includegraphics[width=\textwidth]{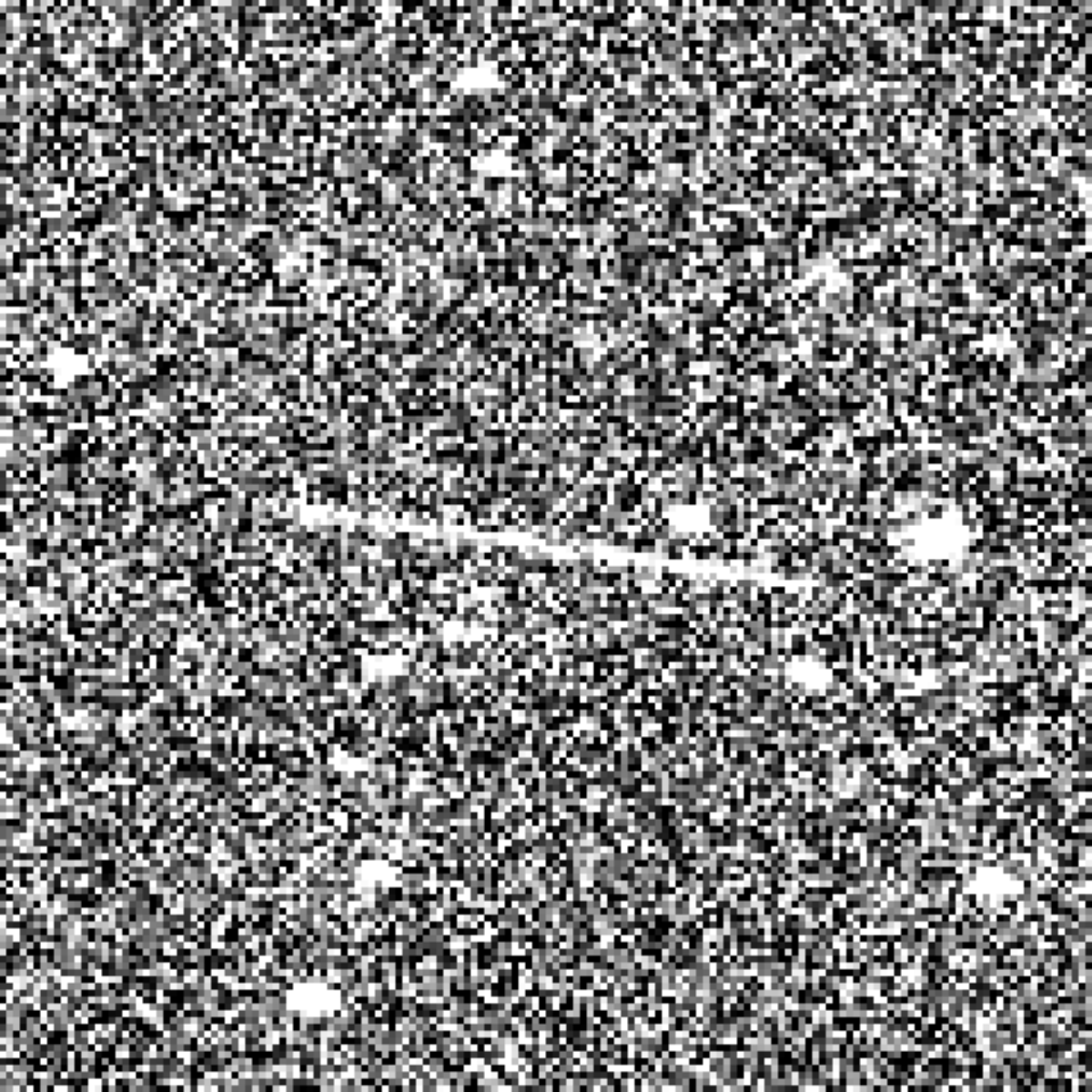}
        \captionsetup{format=hang}
        \caption{0.500\,s exposure\newline binning factor 2\newline 13$^{\rm th}$ mag target}
        \label{fig:example_frame_g}
    \end{subfigure}
    \begin{subfigure}{0.3\columnwidth}
        \includegraphics[width=\textwidth]{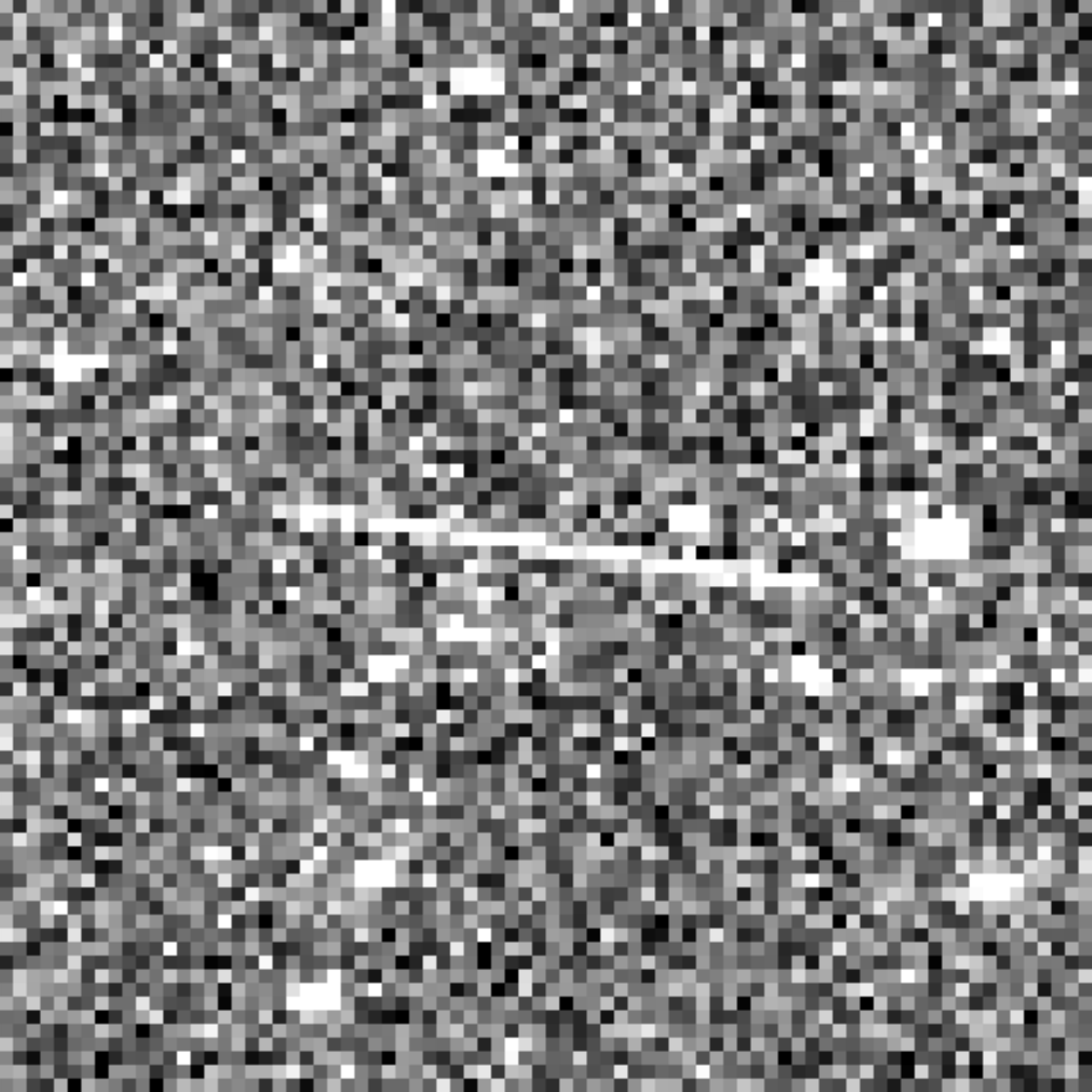}
        \captionsetup{format=hang}
        \caption{0.500\,s exposure\newline binning factor 4\newline 13$^{\rm th}$ mag target}
        \label{fig:example_frame_h}
    \end{subfigure}
    \begin{subfigure}{0.3\columnwidth}
        \includegraphics[width=\textwidth]{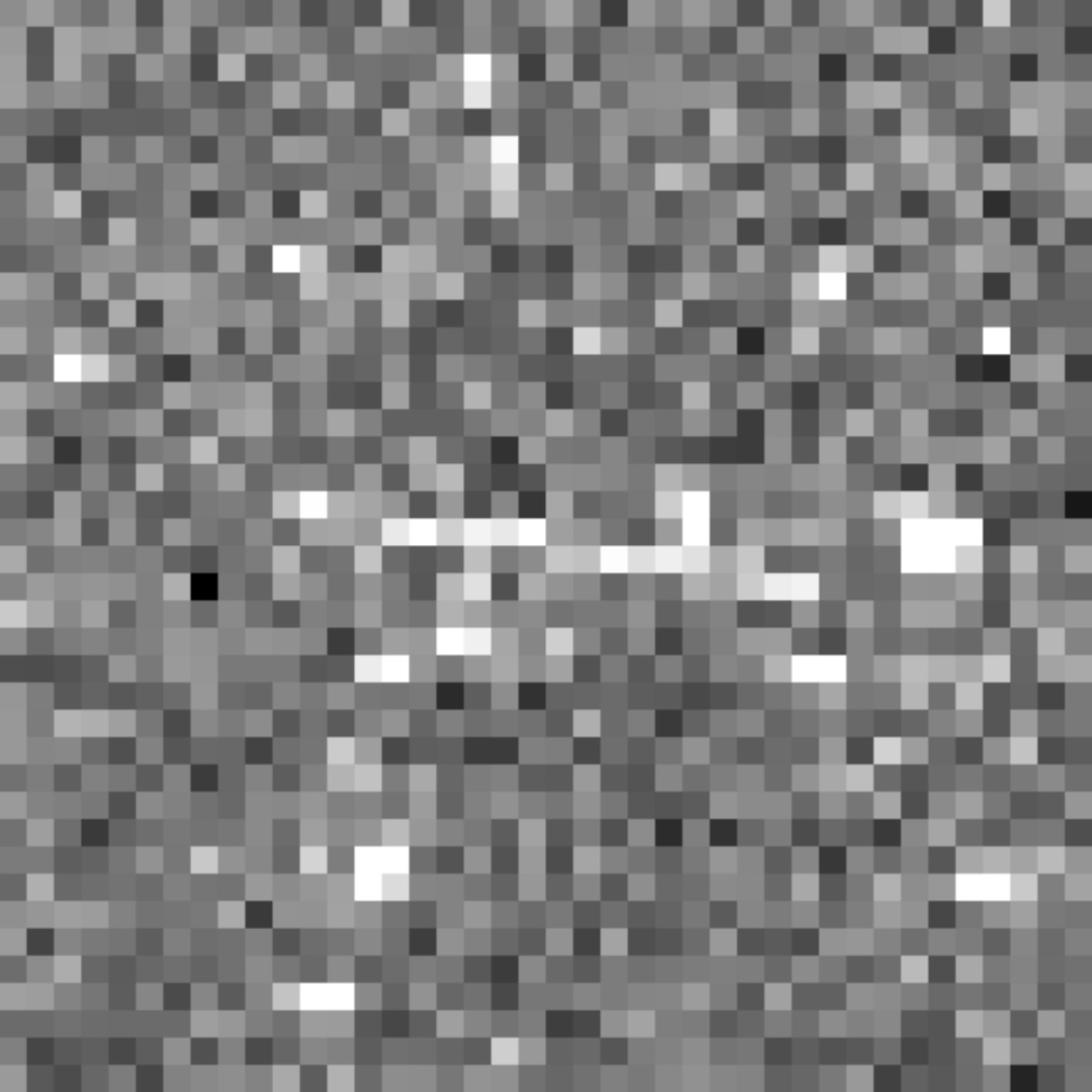}
        \captionsetup{format=hang}
        \caption{0.500\,s exposure\newline binning factor 8\newline 13$^{\rm th}$ mag target}
        \label{fig:example_frame_i}
    \end{subfigure}
    \caption{Example injected signals. Shown for a range of exposure times, binning factors and target magnitudes. Simulated signals are placed within true observational data.}
    \label{fig:example_frames}
\end{figure}

Figure \ref{fig:example_frames} shows some examples of injected signals. In each sub-figure a target of given magnitude is simulated and injected into a real observed data frame. All simulated targets have the same velocity so as to allow for easy comparison. The exposure time and binning factor used for the generation of each simulated target is given. Slower moving, brighter targets are chosen here so as to be visible in individual example frames for demonstration purposes. Additionally, image colour scale and magnification are optimised to most clearly display the injected signals, parameters which cannot be known for random targets. Visible also in these example frames is the result of longer exposures on the star trails. Since the telescope set-up is stationary, stars trail slightly in individual exposures and longer exposures led to longer trails. From these example frames we see that this effect is negligible in individual images, validating the use of the stationary telescope system.

The simulated signals as shown here are simplifications of observed LEO signals which can show significant variability on a range of time scales. The sources of variability can be caused by a range of effects, including the motion of the target, (i.e. tumbling satellites can show brightness variations of many percent), atmospheric effects (i.e. scintillation, which can cause brightness fluctuations of tens of percent) and instrumental effects (i.e. cross-track jitter, caused by the uneven distribution of flux across multiple pixels due to the fact that targets are not being spatially resolved). These myriad effects all result in a target brightness profile that can be variable on short time scales. However, the methods discussed in this paper are all based on average brightness values. The stacking and integration techniques discussed depend on the total brightness observed across a full exposure, or sequence of exposures. Therefore, the changing brightness profile is not the figure of merit, only the average brightness across the relevant observation window. As such, we do not simulate these additional effects, and all results are given in terms of the average brightness of a target.

\subsection{Blind stacking}
\label{sec:Blind stacking}

With the simulated streaks properly injected into the data the next step is the actual blind stacking procedure. Blind stacking works by making assumptions about the motion of a potential target and then testing the data for a target with that motion. This process is repeated for all potential motions that one wishes to consider with the results being combined into a single image. The procedure followed in this paper is broadly similar to that laid out in \citet{2020PSJ.....1...81R}.

\begin{figure}[htp]
    \centering
    \includegraphics[width=0.9\columnwidth]{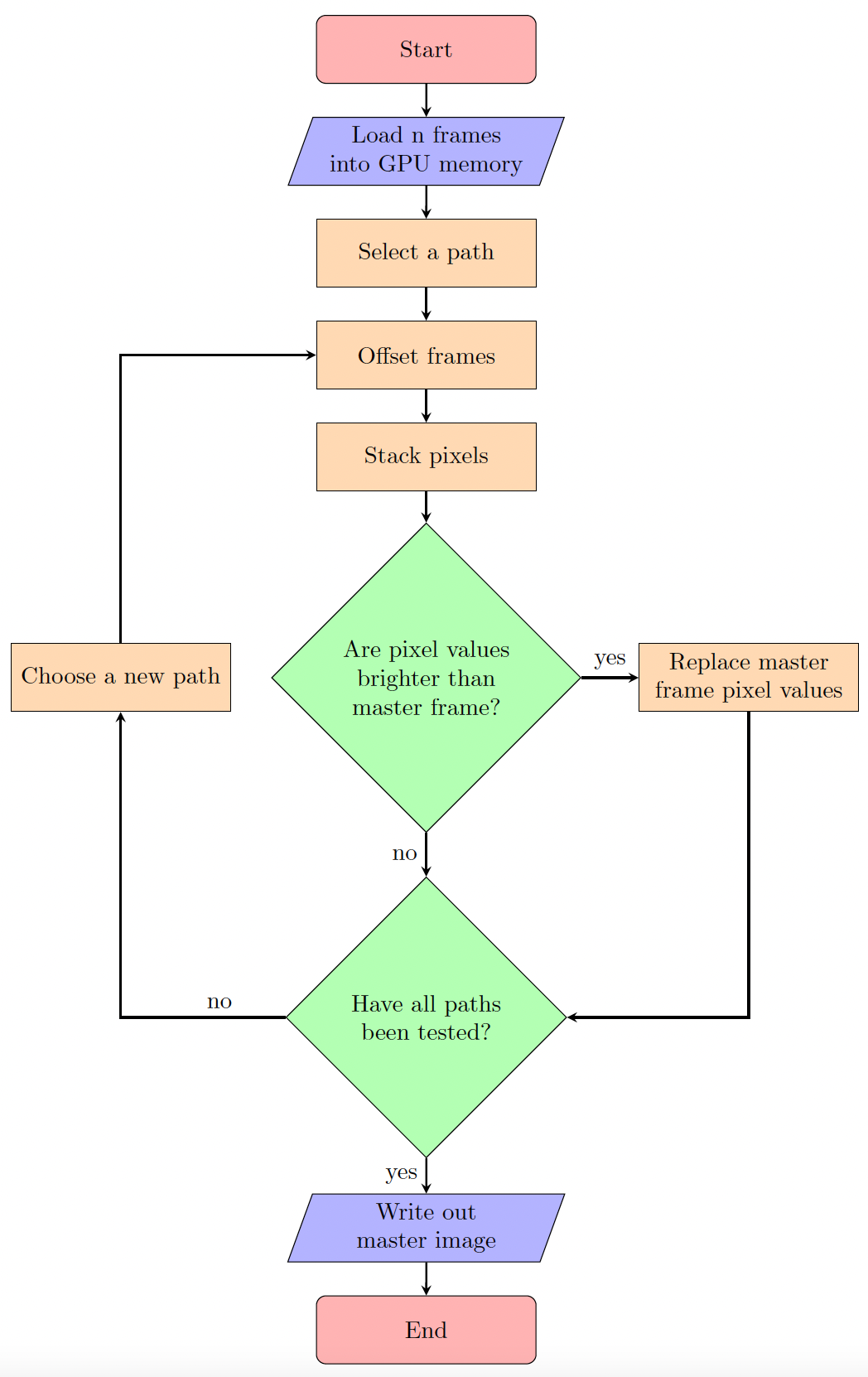}
    \caption{Graphical representation of the blind stacking pipeline.}
    \label{fig:flowchart}
\end{figure}

A potential motion is chosen and decomposed into an $x$ and $y$ component, ($v_x$, $v_y$) measured in number of pixels. At each time-step in the data the corresponding frame is offset from the first by ($iv_x$, $iv_y$) where $i$ is the index of the frame (zero-indexed). This leads to a stack of pixels at each pixel location which are combined to give a single value for each pixel. The method used to combine the stacked pixels is a trimmed mean, in which we reject the brightest and faintest pixels in a stack and then average the surviving values. Only pixels overlapping the first image are considered, pixels whose offsets move them off the footprint of the original frame are considered lost. This process is then repeated for every potential motion to be considered. After each tested motion the stacked images are compared with only the brightest pixel at each location being saved (along with the meta-data about which tested path produced it). This results in a single master stacked image, the same dimensions as the original frame, with a single value at each pixel location, but that comprises information about any potential path. Targets which have moved at a constant rate between frames will align when the correct path is tested, resulting in a brightness spike in the stacked image. These bright spikes are thus indicators that there is a moving object at that location in the original frame. The meta-data at that location then reveals the $x$ and $y$ motion of said object. Figure \ref{fig:flowchart} shows a graphical representation of this process. For a pseudo-code description of the blind stacking algorithm see \ref{app:blind_stacking_pseudocode}.

The number of paths that must be tested is related to the range of velocities a potential target might reasonably be expected to have (see Section \ref{sec:Target velocities}). Because the direction of the object is unknown, testing a particular magnitude of motion requires testing all paths that have that magnitude, regardless of direction. With an upper and lower limit, $v_{\rm max}$ and $v_{\rm min}$, the number of paths that must be tested is roughly given by the following equation,

\begin{equation}
\label{eq:n_paths}
    n_{\rm paths} \approx \pi(v_{\rm max}^2 - v_{\rm min}^2).
\end{equation}

where $v_{\rm max}$ and $v_{\rm min}$ are measured in pixels. The introduction of a lower limit saves little in the number of paths but has the added effect of removing spikes from stars, which are typically moving much slower than any potential LEO target. Preferring specific directions could limit the number of required paths dramatically but would result in a reduction in generality of the technique. If searching for a specific subset of targets however, it would be a useful mitigation.

The presence of noisy and/or star contaminated pixels, combined with the algorithm's process of choosing the brightest stacked pixel at each location, every path tested has the possibility of lining up high noise pixels and creating a spike. This means that the more paths that are tested, the greater the average pixel value in the master stacked image and therefore, the brighter a true spike must be to have the same signal to noise ratio. It is also noted that longer exposures create longer star trails, increasing the number of bright background pixels and exacerbating this issue, though this affect is negligible at the exposure time and binning factors considered here. 


Therefore, limiting the number of paths is vital to the success of the blind stacking technique.

\subsection{Optimisation parameters}
\label{sec:Optimisation parameters}

There are three key physical optimisation parameters that we have considered in this study. These are the number of frames used in each data-set, $n$, the exposure time of each frame, $t$, and the amount of binning carried out before stacking the frames, $b$.

\subsubsection{Number of frames}
\label{sec:Number of frames}

\begin{figure}[htp]
    \centering
    \includegraphics[width=\columnwidth]{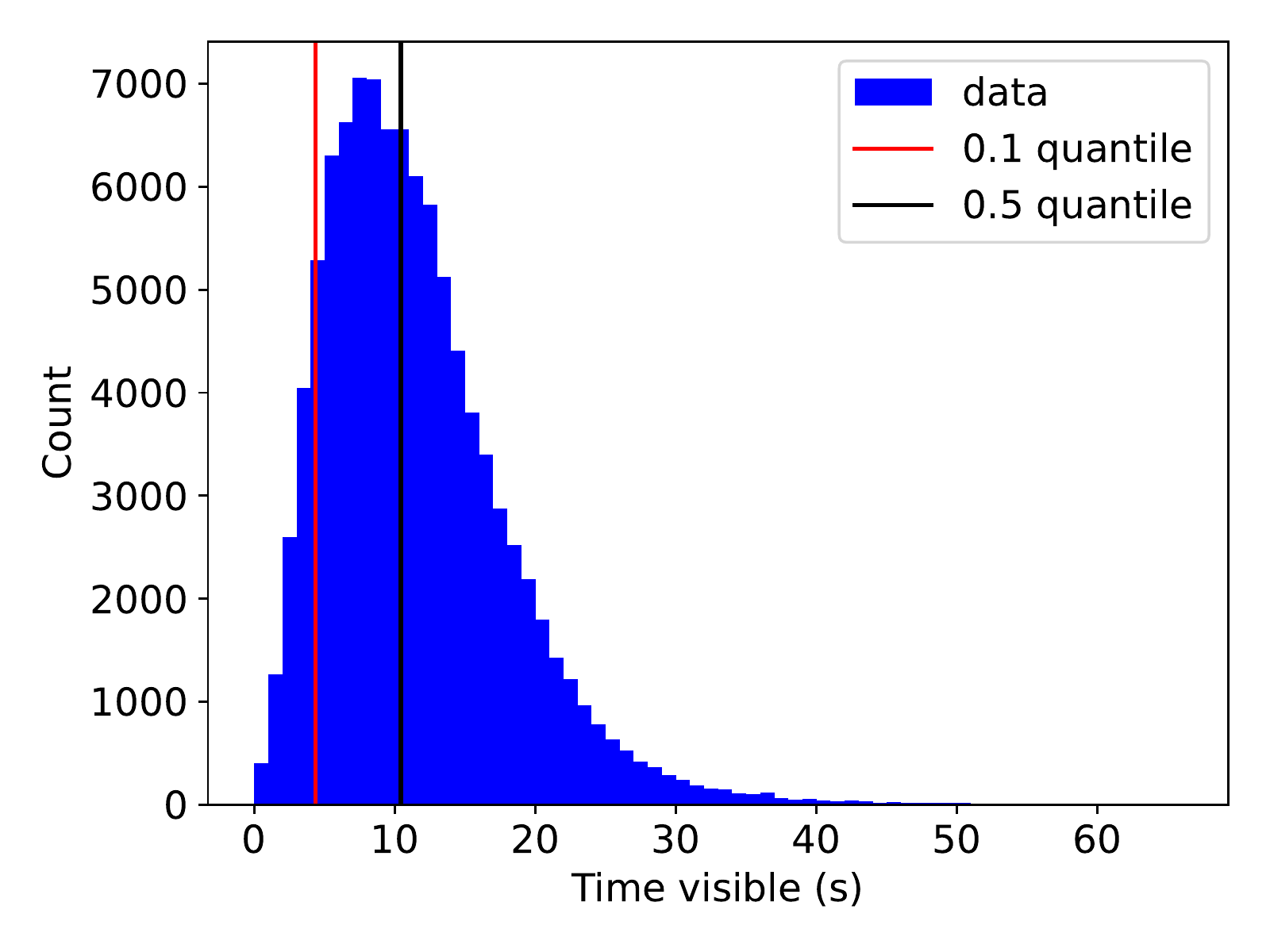}
    \caption{Distribution of the time for which simulated targets are within the field of view. Also included are the 0.1 and 0.5 quantiles.}
    \label{fig:average_time}
\end{figure}

The optimal number of frames per data-set, $n$, is a factor of exposure time and target velocity. The ideal observing time (exposure time $\times$ number of frames) for a specific target should be equal to half the time for which a target will be within the field of view. This maximises the number of frames a target will appear in whilst ensuring the target will appear in the first frame of at least one data-set. Appearing in the first frame is a requirement for an object to be detected since this is the frame against which all path are measured (although this requirement will be further investigated in following work). Observing for longer than this amount of time increases the SNR of objects which appear in all frames (since more frames are stacked together) but reduces the fraction of objects which appear in all frames (since it requires them to remain in the FoV for longer). Observing for a shorter amount of time increases the chance that an object will appear in all frames but reduces the number of frames that can be stacked for each object, reducing the SNR. Simulating multiple objects drawn from the speed distribution shown in Fig. \ref{fig:velocities} and allowing them to cross a random part of the FoV shows that 90\% of targets are within the field of view for $\geq4.3$s (CLASP has a field of view of $2.63 \times 1.76$\,deg). To detect a target it must appear in the first frame of a data-set so observing for $(4.3/2)=2.15$s means that we will have a data-set that has 90\% of targets visible in the first frame. Observing for less than this would result in only a small increase in detections. This corresponds to $\sim17$ frames of 0.125s exposures. Conversely, only 50\% of targets are visible for $\geq10.4$s. Observing for longer than $(10.4/2)=5.2$s would make these slower targets clearer but would both increase memory requirements (more frames must be stored at once) and make it more likely that faster targets would not be caught in the first frame of a data-set. Additionally, these slower targets are easier to detect anyway since their light is spread across fewer pixels in each individual image. This corresponds to $\sim42$ frames of 0.125s exposures. Fig. \ref{fig:average_time} shows this distribution including the 0.1 and 0.5 quantiles.

With these values in mind (but adapting for ease of computation) the $n$ values we choose to test are 24, 40 and 56 for $t=0.125$s; 12, 20 and 28 for $t=0.25$s; and 6, 10 and 14 for $t=0.5$s. These values correspond to total observing times of 3, 5 and 7 seconds.

\subsubsection{Exposure time}
\label{sec:Exposure time}

Decreasing the exposure time, $t$, reduces both the background noise of the individual frames (up to a threshold, see 
Fig. \ref{fig:zp_noise}) and the number of paths that must be tested between successive frames, subsequently reducing the run-time. The exposure times discussed in this report are 0.125s, 0.25s and 0.5s. These exposure times are chosen in order to test a range of parameter space while allowing for hardware and software constraints (see Section \ref{sec:Data acquisition and reduction}). Going to shorter exposures would require the telescope system to run at higher cadences, which is currently not an option with our observational set-up. Longer exposures would result in intractable run-times that would make the simulations and analysis untenable. The chosen exposure times give a good range and clear display of the effect of exposure time on recovery and run-time. 
Longer exposures mean the number of pixels a target can move
between frames increases and therefore the number of testable paths also increases as the maximum motion squared.

\subsubsection{Binning}
\label{sec:Binning2}

Binning refers to the combination of pixels in each individual frame before proceeding with the blind stacking procedure. A binning factor of $b$ means that groups of $b$ by $b$ pixels are averaged together, resulting in $1/b^2$ as many pixels in the binned image than in the unbinned one. Binning is implemented in both the $x$ and $y$ directions with pixels being averaged into single bins. Any edge pixels which cannot form an entire bin are discarded (thus the frame may lose the data from up to $b-1$ pixels in both $x$ and $y$ directions after binning, for a binning factor $b$). Binning is implemented after reduction and background removal and after the injection of any simulated signals.

The binning used in the results discussed here is achieved by taking the mean value of the binned pixels, this is most equivalent to the real world case of scaling up the pixel size of the telescope (although it doesn't account for a changing field of view). The chosen binning values to be tested in these results are 2, 4 and 8. We don't test the unbinned images since, firstly, this would require a significant run-time, and secondly, the individual unbinned images contain a subtle but measurable odd/even pixel offset such that odd/even pixels are drawn from slightly different distributions. Binning by an even value removes the effect. Additionally, binning by a power of 2 is easiest computationally.

As well as reducing the number of pixels per image, binning can also potentially reduce the number of tested paths by the same factor, however, this then reduces the resolution of the tested paths and thus the recovery fraction. For example, if a target moves 5 pixels per frame in the $x$ direction and the pixels are then binned by a factor of 2 the closest integer-pixel paths that can be tested are 4 or 6. After 40 frames the target will have moved 200 unbinned pixels whereas the closest paths will have moved 160 or 240 unbinned pixels, thus they will be stacking the wrong pixel values. The larger the number of frames and the larger the binning factor the worse this effect can become. To account for this, while still gaining a benefit from binning the images, we must test non-integer motions in the binned images. As before, consider a path of 5 unbinned pixels per frame. In the unbinned image the position of this pixel is 0, 5, 10, 15, 20... whereas in the binned image the location of the binned pixel that contains this unbinned pixel is 0, 2, 5, 7, 10... We see that this sequence is no longer an arithmetic progression, therefore, to accurately portray this motion we must test a path that alternates between moving 2 and 3 pixels per frame. The result is that, to ensure sufficient resolution in the tested paths, we must test as many paths in the binned images as we do in the unbinned images.

\subsection{Detection code}
\label{sec:Detection code}


Once the simulated targets have undergone blind stacking the resulting images must be searched automatically, and any detections compared to the injected signals. Additionally, we need to search the individual images in a similar manner, in order to quantify the improvement in detectability produced by the blind stacking method. When searching for targets in images we can either search the image as it appears, looking for clusters of bright pixels, indicating an extended object, or we can attempt to integrate along a potential streak, increasing the signal to noise of a detection. Since we don’t know the location or velocity of any target that may potentially be in an image, we need to integrate along all possible paths, in much the same way as we test all potential paths during the blind stacking phase. This is repeated for every individual pixel in the image. We first choose a path from the list of possible paths. At each pixel location we determine which other pixels would contain parts of an extended object, if it were moving along the path being tested and was centred on the pixel being tested. We then calculate the SNR of these pixels using the following equation;

\begin{equation}
\label{eq:snr1}
	{\rm SNR} = \frac{s n}{\sqrt{s n + n \sigma^2}},
\end{equation}

where $s$ is the average signal per pixel, $n$ is the number of pixels, and $\sigma$ is the noise per pixel (assuming negligible read noise from the sCMOS detector). This value is then assigned to the relevant pixel. The same process is repeated for every pixel, generating an SNR image. This is compared to the master image, and only the brightest value at each pixel location is kept. We repeat these steps for every potential path. The result is an image in which the pixel values correspond, not to brightness, but to the maximum SNR of an pseudo-aperture placed at that location, with parameters corresponding to a potential path. For a pseudo-code description of the integration algorithm see \ref{app:integration_pseudocode}.

We can then search this image for detections using SEP. This allows us to detect fainter targets than searching the pre-integrated image, since we are considering the integrated flux, as opposed to the flux of individual pixels. Using the above SNR equation, the SNR of a streak of $n$ pixels is $\sqrt{n}$ times larger than the SNR of an individual pixel. However, it must also be noted that, since we must test every potential path, any individual pixel has $n_{\rm paths}$ distinct SNR values generated, and only the largest value is saved. This means that the more paths that are tested, the brighter the background level in the SNR image, while the integrated flux of the true detection is unchanged. When searching for detections it is the difference between the target SNR and the background SNR level that is the relevant quantity. Because of this effect, the SNR of true detections decreases with the number of tested paths and improvements in detectability resulting from integration are less than would be expected in the traditional case where the target location or direction of motion is known. Dividing this quantity by the noise in the integrated image gives the number of standard deviations of a detection above the background level. To limit the number of false positives we set the required threshold to 5 standard deviations above the mean.

One additional aspect that must be considered when integrating the individual images is that the stars must first be masked out. Since the stars are bright and occupy more than a single pixel (due to the telescope PSF) the integration step may highlight short paths which cross over a bright star. Masking these stars using SEP first, reduces the false positives significantly. This is not necessary when integrating the stacked image since the stars are not seen due to the choice of allowed paths.

To compare detections in the stacked image to those in the individual images we must perform the same integration step. The integration procedure used to produce the results seen in this paper is performed only on the final result of the blind stacking algorithm. This is computationally simple because the stacking pipeline (which should ideally, be close to real time) is separated from the integration and detection pipeline (which can occur later and need not be real time). For a more complete integration analysis the integration step can be run in tandem with the blind stacking process. After each path-specific master frame is generated, the integration step is implemented, using just the path being currently tested, before the result is then compared against the final master image (as seen in the flowchart above). This has the benefit that the integration step considers only the pixels generated by stacking along the same path, whereas the previous method can integrate along pixels generated by stacking multiple different paths. The downside is that the integration step is now part of the blind stacking pipeline and therefore, must ideally be run in real time. In practise, this full method is too computationally intensive to justify the minor increase in recoverability it produces.



Once the integration steps are done, the resulting images are searched for detections using SEP. The SEP thresholds are chosen to maximise the recoverability of targets, while keeping false positives to a minimum. Cuts are made based on SNR, as well as target shape and size (adjusted by the relevant optimisation parameters). The resulting integrated images show the same extended detections as in the non-integrated frames. The centre of a streak is the brightest and corresponds to testing the correct path at the correct pixel. Surrounding pixels are then fainter, but still brighter than the background level, and correspond to integrating along slightly incorrect paths that contain only part of the true signal, or paths that contain too many background pixels along with the true signal. The location, orientation and length of the detections are compared to those of the injected signals. Detections with parameters sufficiently close to those of a simulated target (allowing for inaccuracies due to binning etc.) are defined as true detections, whereas clusters not matching any of the targets are false positives. Targets which don't have a corresponding detection are defined as not recovered.

\section{Results}
\label{sec:Results}

There are multiple ways in which to evaluate the success of the blind stacking method as a function of the optimisation parameters. The key aspects discussed here are recoverability (the fraction of injected signals that are successfully recovered) and run-time (the time required for the blind stacking code to complete). We can look at how these measurements are affected by the choice of optimisation parameters to determine the optimal combination. Analysed parameters are the number of frames per data-set, $n$, the binning factor, $b$ and the exposure time $t$. Results presented below are averaged across 50 runs, with each run containing 10 unique and randomly generated injected signals. Recoverability is presented as a fraction of potentially recoverable objects (i.e. objects moving too quickly or appearing in too few frames are not included).

\subsection{Recoverability}
\label{sec:Recoverability}

\begin{figure}[htp]
    \centering
    \includegraphics[width=\columnwidth]{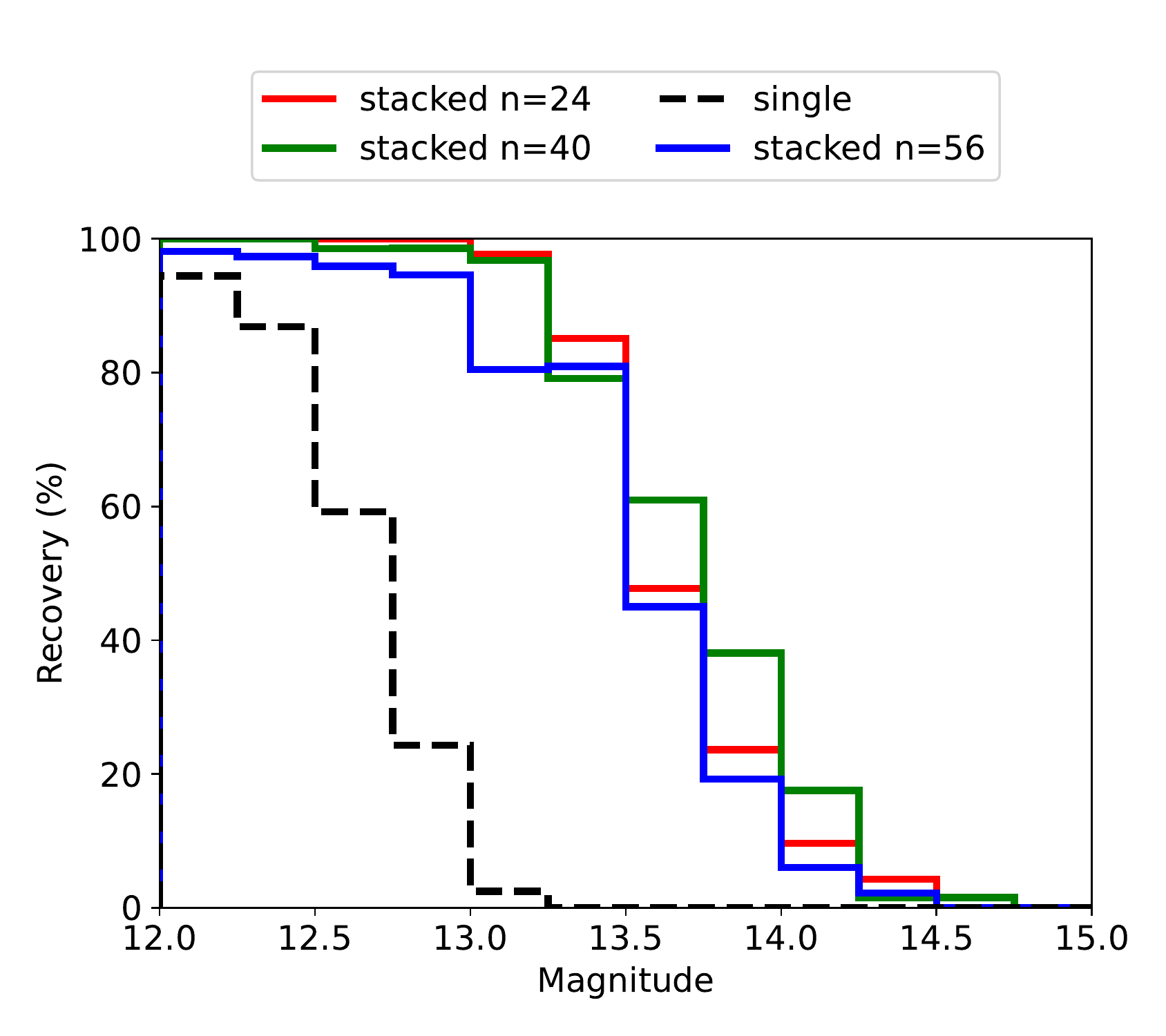}
    \caption{Recovery fraction as a function of target magnitude. Distributions all have an exposure time of 0.125s and a binning factor of 4. $n$=24, 40, 56 results are in red, green and blue respectively. Solid lines show the stacked frame recoverability and the black dashed line shows the single frame recoverability for comparison.}
    \label{fig:nframes_recovery}
\end{figure}

To look at recoverability we look at the output of the detection code and determine what fraction of injected signals are correctly recovered. This can be left as a single value for any combination of optimisation parameters, or, more usefully, broken down by target magnitude or velocity. To best explore how each optimisation parameter affects recoverability we examine each in turn while holding the other two constant (or, in the case of number of frames, relatively constant). We will then plot the results as a function of target magnitude.
Fig. \ref{fig:nframes_recovery} shows the effect of the number of frames per data-set, $n$, on recoverability. In this plot the 3 distributions show the results of using 24, 40 and 56 frames. All distributions use an exposure time of 0.125s and a binning factor of 4.

In this figure we see the impact of the number of frames in a data-set on stacked frame recoverability (single frame recoverability is obviously independent of number of frames in a stack. For the brighter targets we see that the number of frames has a limited effect on the recoverable fraction, with most targets being recovered successfully. This is due to the fact that these brighter targets are more easily identified, sometimes even visible in individual images, thus requiring the stacking of fewer frames to reach the required SNR for detection. Should we test data-sets with many fewer frames we would expect to see a greater decline. At the fainter end of the distribution the number of frames has a more appreciable effect on recoverability. Targets fainter than $\sim13.5^{\rm th}$ mag show a preference for data-sets of 40 frames. Fewer frames than this means the increase in SNR gained when stacking images is not as large as it could be, leading to the loss of the faintest targets. More frames than this creates a longer observing period and therefore means that targets appearing in few frames are less likely to be caught in the preceding data-set (since they would need to appear in the first frame of the preceding data set which is now further in the past). This means targets must now be detected based upon fewer frames, meaning their SNR is reduced and therefore so is their recoverability. 40 frames (or 5s of observing to generalise to all exposure times) is currently the optimal length of data-set.

\begin{figure}[htp]
    \centering
    \includegraphics[width=\columnwidth]{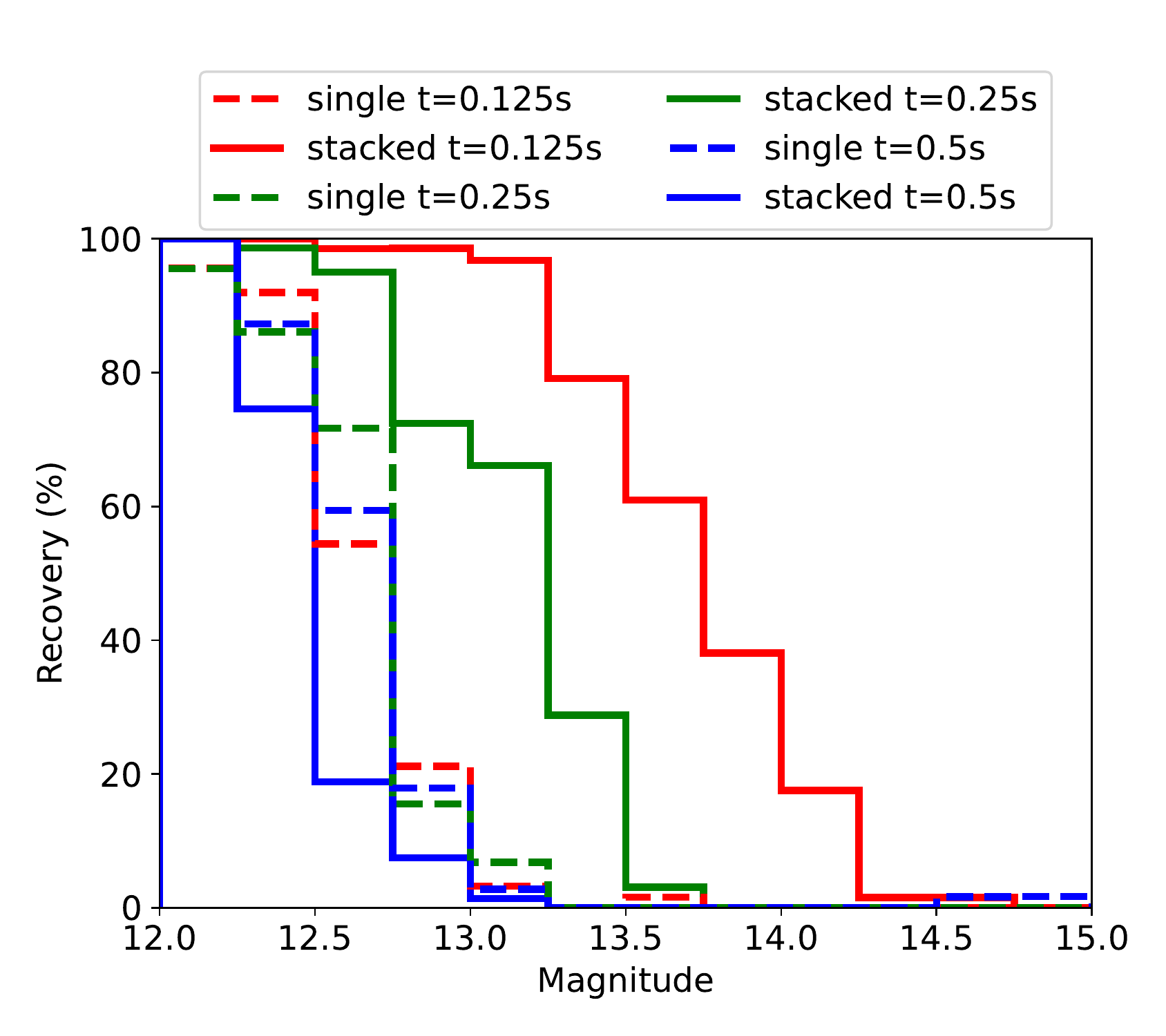}
    \caption{Recovery fraction as a function of target magnitude. Distributions all have an observing time of 5s and a binning factor of 4. $t$=0.125, 0.25, 0.5s results are in red, green and blue respectively. Dashed lines show the single frame recoverability and solid lines show the stacked frame recoverability.}
    \label{fig:exptime_recovery}
\end{figure}

Fig. \ref{fig:exptime_recovery} shows the effect of exposure time. In this plot each distribution has a binning factor of 4 and a total exposure time of 5s (meaning 40, 20 and 10 frames for the different exposure times). The results of using exposure times of 0.125s, 0.25s and 0.5s are seen.

This figure shows the impact of exposure time on target recoverability. For single frame efforts, exposure time has little effect on recoverability. Increased exposure time should lead to a longer streaks without reducing the average flux per pixel, this in turn should lead to an increased SNR as given by equation \ref{eq:snr1}. However, two factors balance this out. The first is the increased noise in individual frames (see Fig. \ref{fig:zp_noise}), reducing the ease of detection. The second effect is that searching for longer streaks necessitates testing more paths. This in turn leads to an increased background level in the integrated image (as discussed in section \ref{sec:Detection code}) meaning that true detections are closer to this limit. These effects combined, lead to a small impact of exposure time on single frame recoverability. The effect on stacked frame recoverability is much more significant, with a strong reduction in recoverability for longer exposures. This is due to similar effects as for the single frame, except now, the increased number of paths affect the results during both the stacking procedure, and the integration stage. This results in a significant degradation of the target signals, so much so that, at $t=0.5$s, the single frame recoverability is superior to the stacked frame recoverability.

\begin{figure}[htp]
    \centering
    \includegraphics[width=\columnwidth]{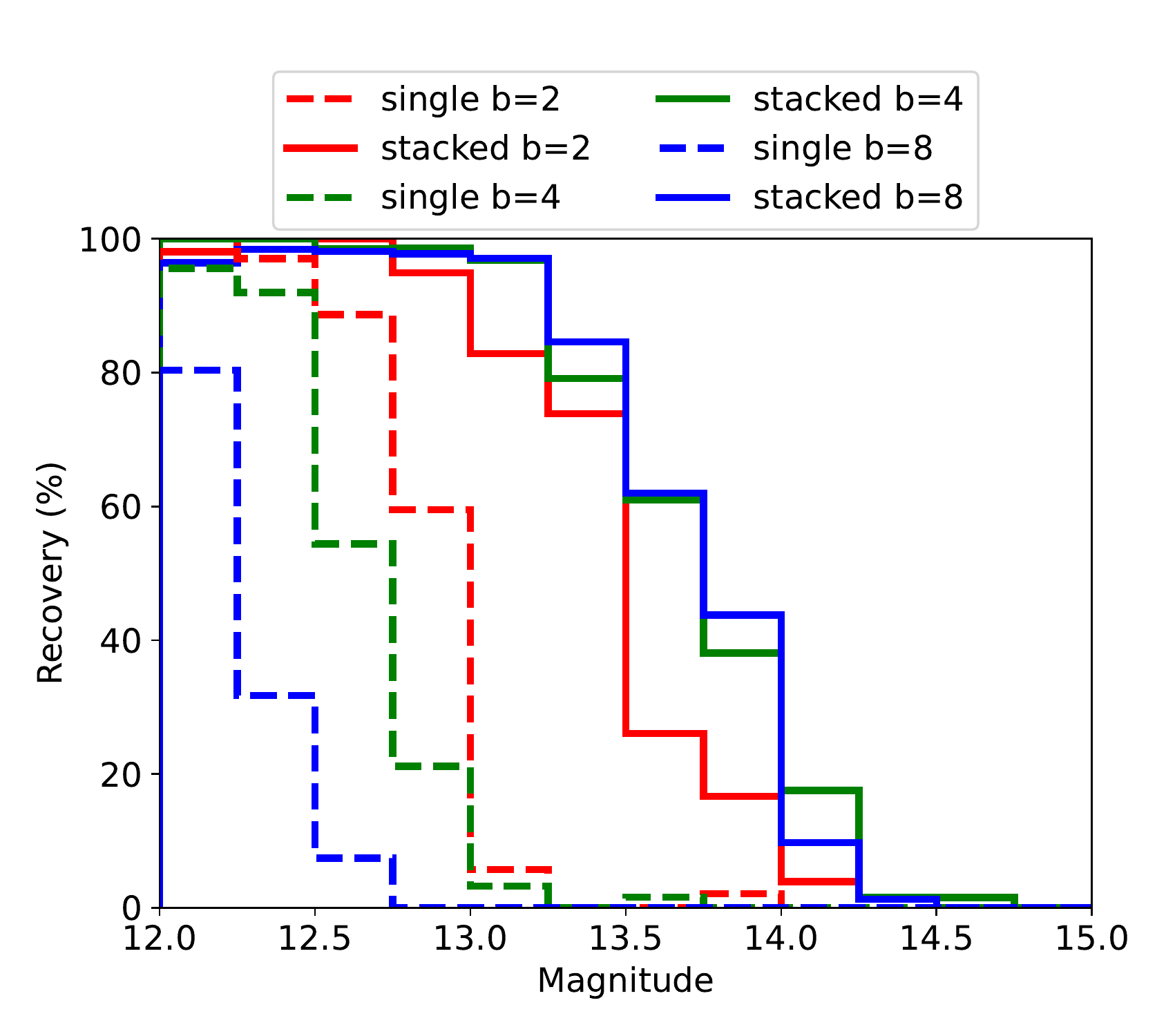}
    \caption{Recovery fraction as a function of binning factor. Distributions all have an exposure time of 0.125s and use 40 frames. $b$=2,4,8 results are in red, green and blue respectively. Dashed lines show the single frame recoverability and solid lines show the stacked frame recoverability.}
    \label{fig:binning_recovery}
\end{figure}

Fig. \ref{fig:binning_recovery} shows the effect of binning factor. Each distribution is run using an exposure time of 0.125s and uses 40 frames. The different binning factors tested are 2, 4 and 8.

In this figure we see how the degree of pixel binning affects the recoverability of targets as a function of magnitude, for both the single and stacked frames. Binning should improve detection up to a limit, since binning pixels reduces the background noise, while concentrating more signal onto fewer pixels. Too little binning means that we do not benefit fully from the noise reduction, whereas too much binning means that the target pixels contain significant background noise. Therefore we would expect recoverability to peak at a binning factor ~4, based on the system PSF.

When looking at the single frame results however, we see that increasing the binning leads to a reduction in the recovery fraction. This is due to inaccuracies during the integration step caused by the loss of resolution when binning. Examining the recovery before integrating we see that the recovery peaks at a binning factor of between 4 and 8, which matches expectations. The integration step however, determines which pixels should contain signal based on the binned image. Increased binning makes these pixels less accurate (i.e. they may not be the ones that contain the maximum signal) leading to a reduced SNR and worse recoverability. For the stacked frames however, the recoverability peaks at a binning value between 4 and 8, as expected. In this case, the inaccuracies introduced during the integration step are offset by the boost in signal resulting from the blind stacking procedure. It therefore turns out that the optimal binning value for recoverability using the blind stacking method is not equivalent to the optimal binning value for recoverability in individual frames.

\begin{figure}[htp]
    \centering
    \includegraphics[width=\columnwidth]{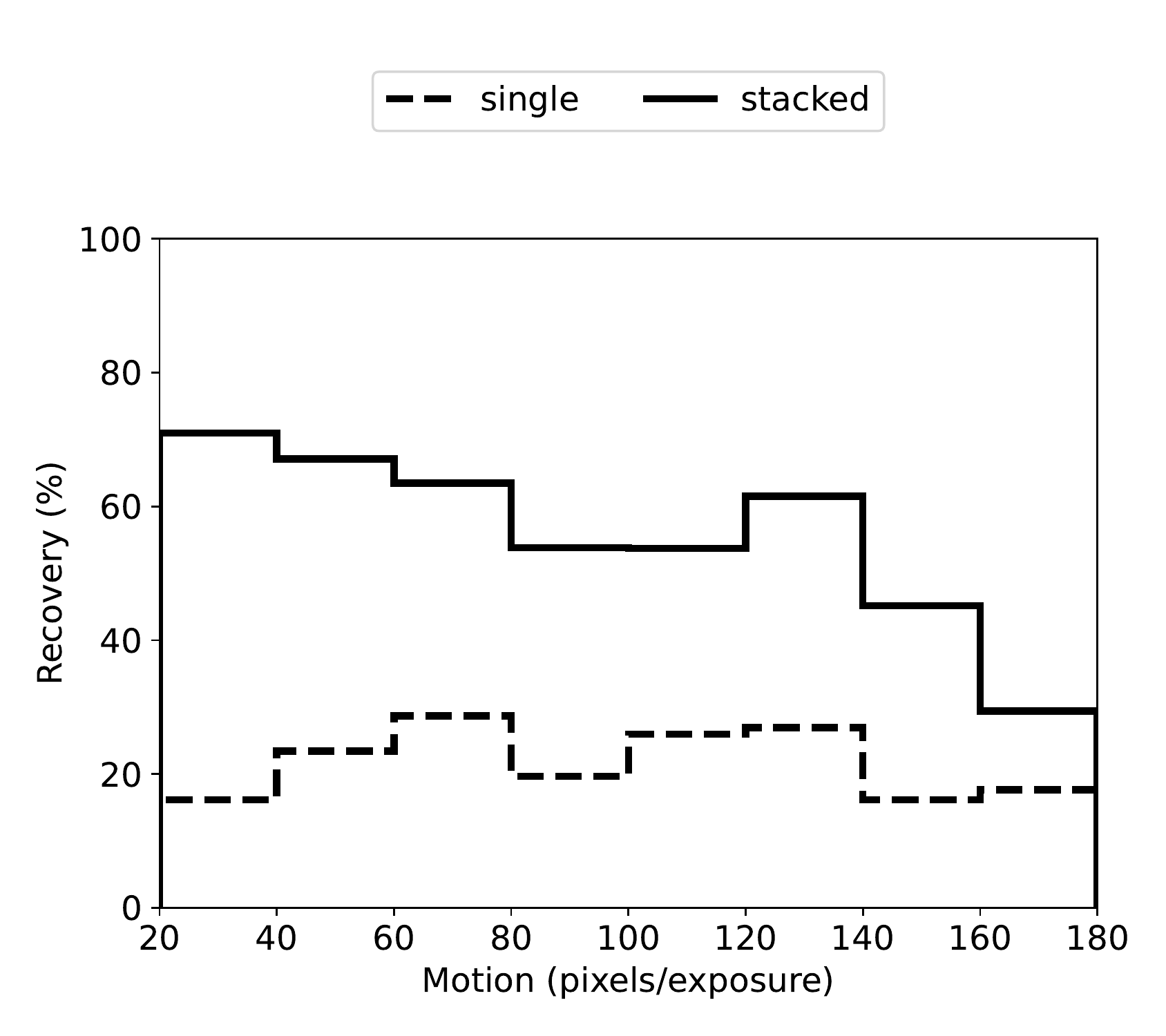}
    \caption{Recovery fraction as a function of target speed. Distributions have an exposure time of 0.125s, a binning factor of 4 and use 40 frames. The dashed line shows the single frame recoverability and the solid line shows the stacked frame recoverability.}
    \label{fig:motion_recovery}
\end{figure}

Alongside looking at recoverability as a function of magnitude we can also look at the effect of target velocity. 
Fig. \ref{fig:motion_recovery} shows the recoverability 
as a function of target velocity. This run uses an exposure time of 0.125s, a binning factor of 4, and 40 frames.


In this figure we see that for a single frame, recoverability is broadly independent of target velocity, whereas in the stacked frame recoverability falls with increasing target velocity. This is due to the combination of two distinct regimes. In the bright regime, simulated targets are at or above the imposed pixel threshold value used to reduce the false positive effects of bright stars. For these sufficiently bright targets, increasing the speed increases the number of pixels over which the signal is spread, but does not significantly decease the signal per pixel. Therefore, the total signal is seen to increase and the SNR rises. For fainter targets (the majority of targets) this threshold is not an issue and the signal per pixel falls as the number of pixels rises with increasing target velocity, leading to a reduction in SNR. When looking at single frames, few faint targets are recovered, therefore these two effects cancel out and lead to an apparent independence of recoverability with target velocity. In the stacked frames however, more faint targets are seen so the faint target behaviour is enhanced leading to the displayed inverse relation between target velocity and recoverability.


\subsection{Predicted recoverability}

\begin{figure}[htp]
    \centering
    \includegraphics[width=\columnwidth]{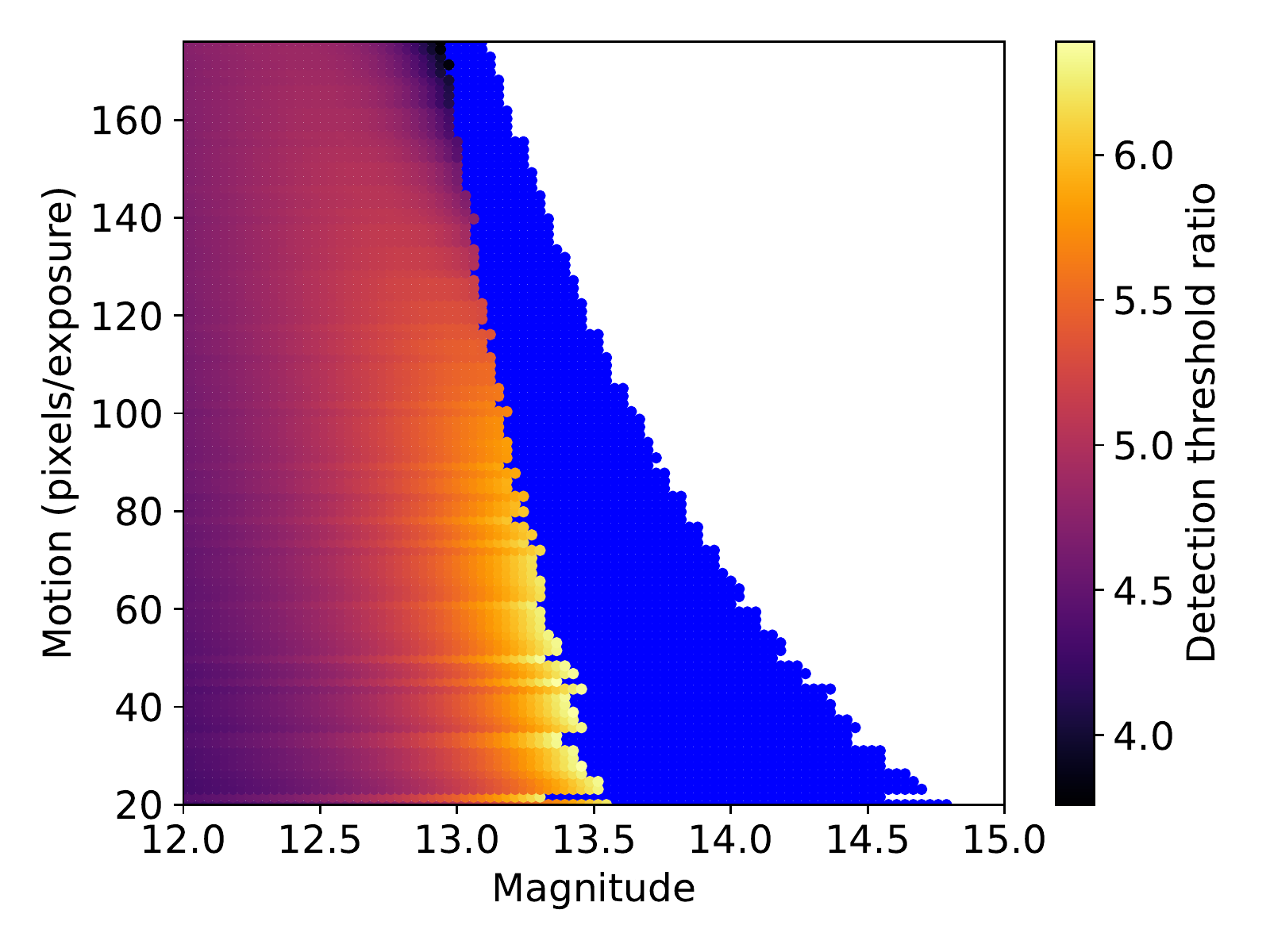}
    \caption{Recoverability as a function of target magnitude and speed. Points accessible to both the single and stacked frames are coloured by the ratio of the detection threshold in both cases, with positive values meaning the target is more easily detected in the stacked frame. Targets only accessible in the stacked frame are coloured blue. There are no targets accessible only in the single frames.}
    \label{fig:recoverability_limits}
\end{figure}

We have seen how the recoverability of targets is affected by the optimisation parameters in our simulation but it is important to consider how this compares with what we predict the detectability limits to be. To do this, we use the frame background levels and their respective noise values, combined with the results of Equation \ref{eq:snr1}. For a given combination of target magnitude and velocity we can then predict the number of standard deviations above the noise a detection would appear in both an integrated single image and an integrated stacked image. This predicted recoverability can then be compared to the results of our simulations to determine how close to the ideal case our results are. In the predicted results we assume a given target is moving a track oriented at 45 degrees to the positive $x$ axis of the images. A track of this direction means that the pixels chosen when calculating the SNR will be optimal to collect the maximum amount of signal while limiting the background. Additionally, the binning is carried out centred on the target, meaning that the central track of binned pixels will contain the maximum amount of signal. In the full simulation, targets can be injected moving in any direction and at any location. This means that, in practise, the combination of the binning and selection of the streak pixels may be sub-optimal. Therefore we would expect the simulated data to perform less well than the predicted results, although patterns in the data should be replicated. Figure \ref{fig:recoverability_limits} show the predicted recoverability limits for a simulation using $t=0.125$s, $n=40$ frames and $b=4$.

This figure uses the same required threshold of 5 standard deviations above the mean for a detection. Targets that reach this threshold in both the single frame results and the stacked frame results are coloured by the ratio of the two thresholds, with positive values meaning a target is detected more easily in the stacked image. Targets which only reach the required threshold in the stacked image are coloured blue. There are no targets which only reach the required threshold in the single images. From this figure we see that, for this combination of optimisation parameters, we should be able to detect targets in the 13-13.5 magnitude range in single integrated frames, with a preference for slower moving targets. Moving to the stacked frames however, we expect detections of targets up to a magnitude of 14.5, again, with a preference for the slower moving targets. These predictions match well with the results seen in Figure \ref{fig:exptime_recovery}. The recoverability seen in the the simulated data is slightly lower than predicted here, but that is expected as mentioned above. The same predictions can be made for any combination of optimisation parameters but each show good agreement with the simulation results.

It is notable that the improved detection limits in the stacked case are not simply equal to the detection limits of a single frame multiplied by $\sqrt{n}$ as would be the case in traditional stacking. As mentioned above, each incorrect path tested increases the background level of the stacked image, while the brightness of the true detection remains constant, regardless of the number of paths tested. Therefore, the relevant value, i.e. the signal above the background level, decreases with number of paths tested, resulting in a reduced stacking improvement when compared to traditional stacking methods where only one path is tested.

\subsection{Run-time}
\label{sec:Run-time}

\begin{figure}[htp]
    \centering
    \begin{subfigure}{\columnwidth}
        \centering
        \includegraphics[width=\textwidth]{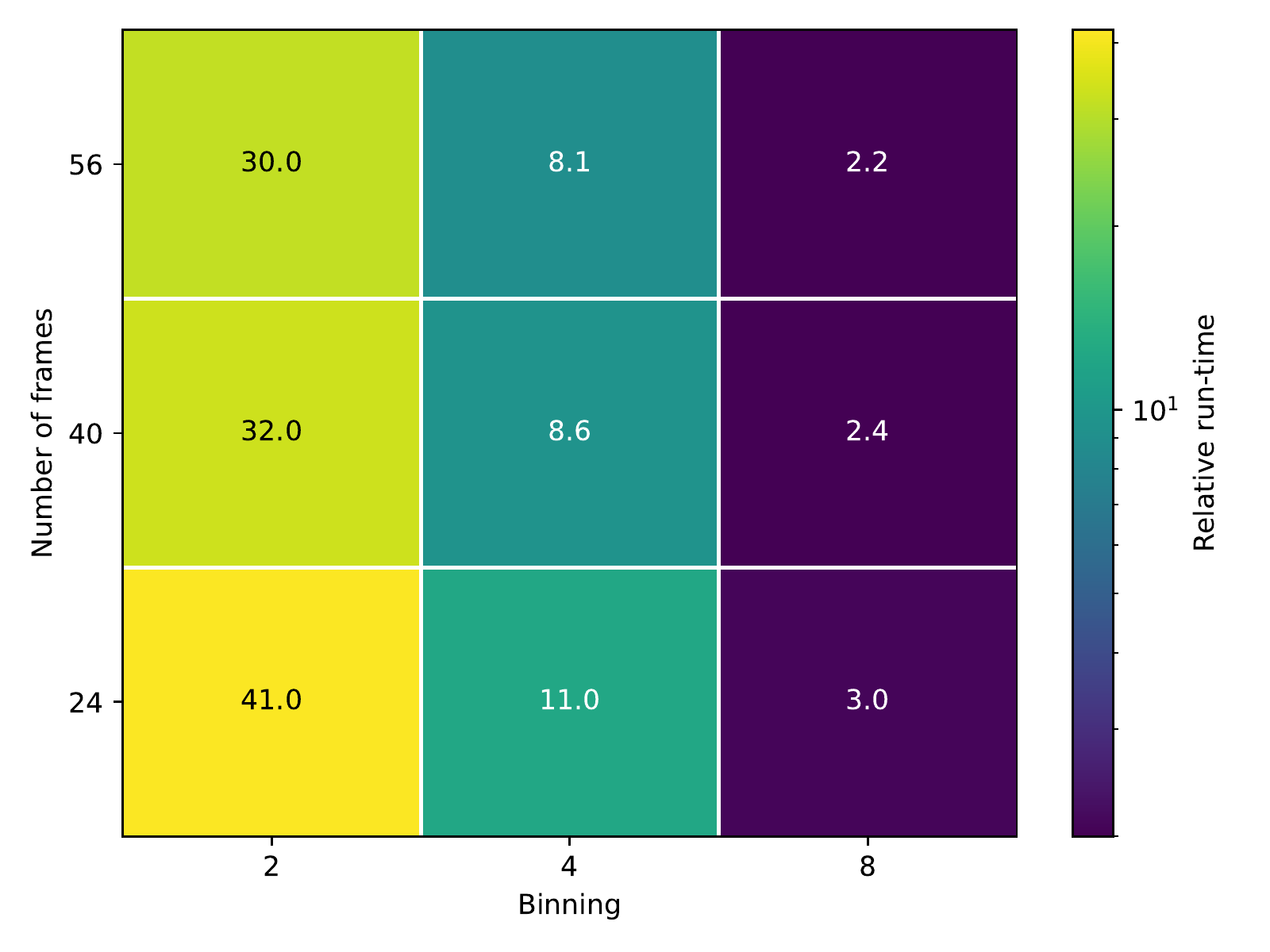}
        \caption{exposure time 0.125s}
        \label{fig:time_125}
    \end{subfigure}
    \begin{subfigure}{\columnwidth}
        \centering
        \includegraphics[width=\textwidth]{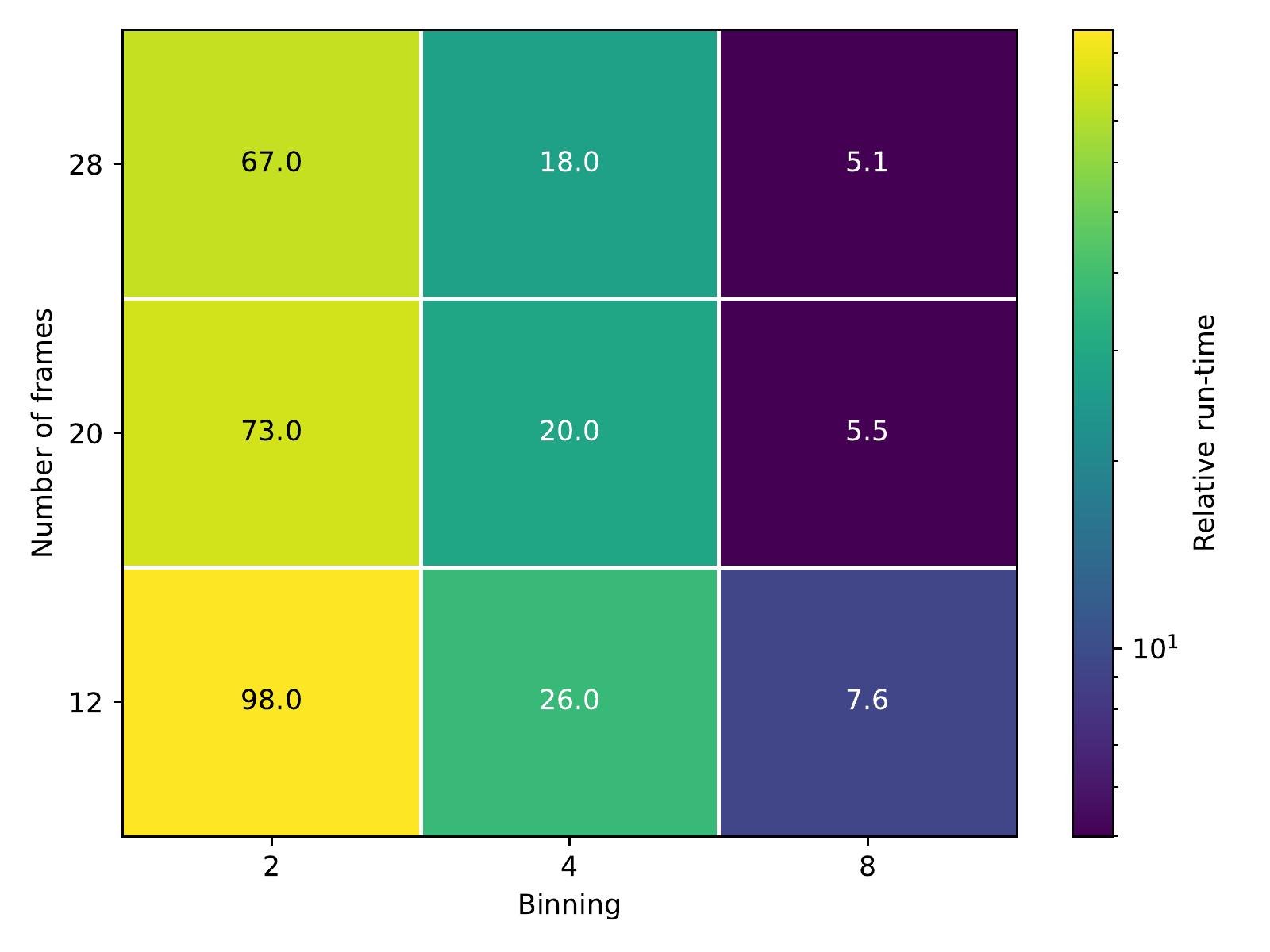}
        \caption{exposure time 0.250s}
        \label{fig:time_250}
    \end{subfigure}
    \begin{subfigure}{\columnwidth}
        \centering
        \includegraphics[width=\textwidth]{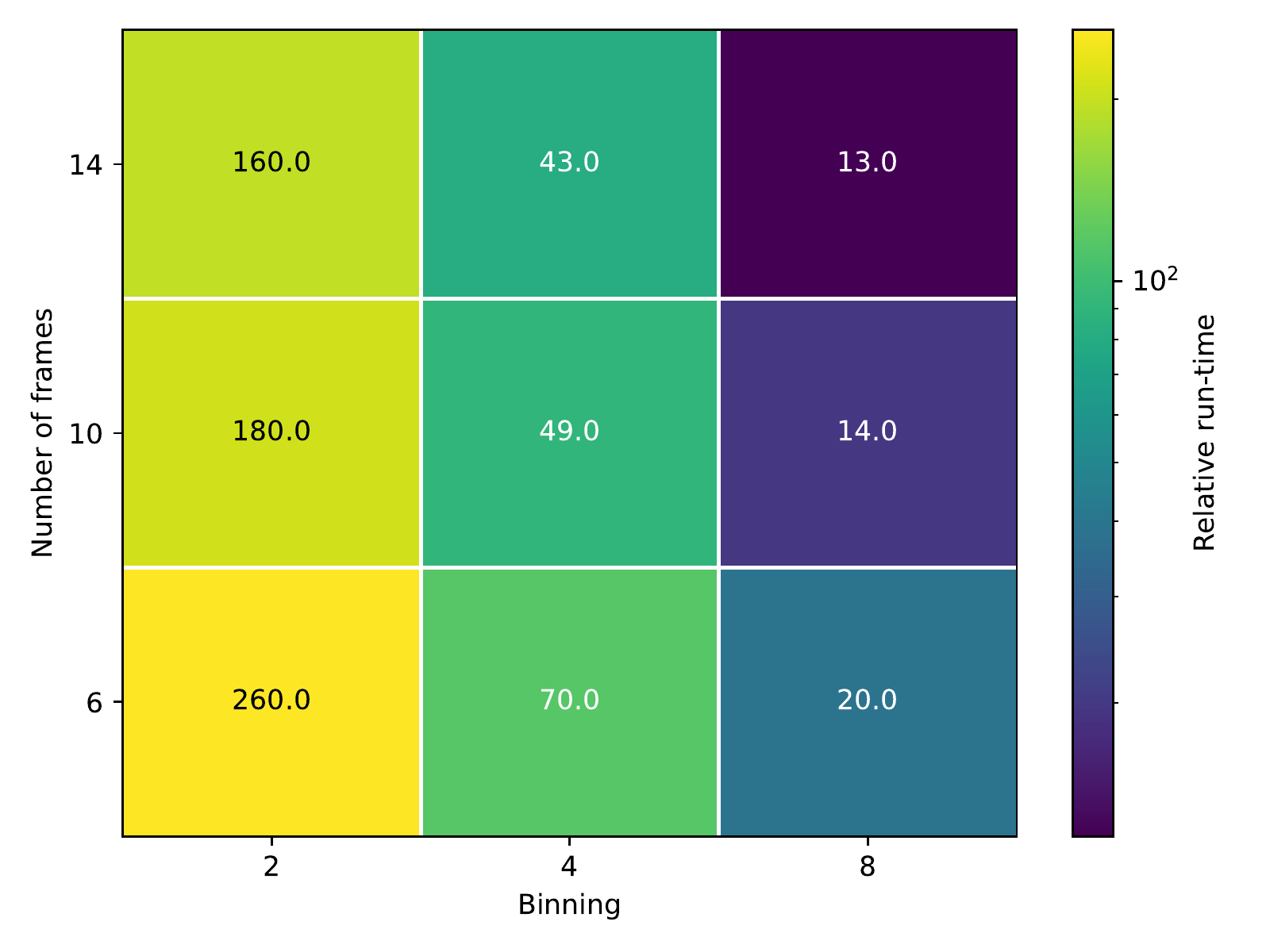}
        \caption{exposure time 0.500s}
        \label{fig:time_500}
    \end{subfigure}
    \caption{Relative run-time (blind stacking algorithm runtime divided by observing time) for all combinations of optimisation parameters.}
    \label{fig:time_all}
\end{figure}

Finally we look at the impact of optimisation parameters on algorithm run-time. The results presented here show the run-time of the blind stacking algorithm divided by the total observing time of a data set (i.e. $n \times t$). Values greater than/less than one mean the algorithm takes more/less time to run than the time used to collect the observations. Fig. \ref{fig:time_all} shows how the algorithm run-time changes with the combination of optimisation parameters used.

From this figure, we see how the algorithm run-time divided by observing time, or relative run-time, changes with each optimisation parameter individually as well as with each parameter combination tested. Firstly, we notice that the relative run-time increases with exposure time. This is due to the fact that we must test more individual paths between frames when the exposure time is longer. 
Longer exposures would also mean longer observing times, but we adjust the number of frames accordingly. Thus the shortest relative run-time corresponds to the runs with the shortest exposure time. Secondly, we see that relative run-time decreases with binning factor. Binning factor affects run-time by reducing the number of pixels which must be analysed. The number of pixels in the binned image is reduced by a factor of $b^2$ which reduces the run-time by the same factor. Binning can also affect the number of paths tested, as discussed in Section \ref{sec:Binning2}. For the method used here, however, the number of paths is unchanged compared to the unbinned approach. Finally, we see the effect of the number of frames per data-set. This produces the smallest effect since, although run-time increases with number of frames, so does observing time. There is, however, a small decrease in relative run-time for a larger number of frames. This is due to the fact that the more frames that are included in a data-set, the greater the number of paths that don't reach the limiting number of frames. Therefore, longer sequences of frames can avoid testing slightly more paths, reducing the relative run-time and resulting in the observed pattern. Based on the results from this figure, and the desire for a real-time or close to real-time system, the preferred combinations of optimisation parameters would be 0.125s exposures with a binning factor of 4 or 8, or 0.25s exposures with a binning factor of 8. Other combinations currently require relative run-times that are unrealistic.

Results presented here have have been computed using the NVIDIA A30 GPU for the bulk of the processing. GPU processing allows for a significant speed up over CPU because large amounts of the algorithm can be parallelised. With the exception of edge cases, all pixels in an image are treated identically during the blind stacking pipeline, additionally the resulting stacked value for a given pixel is not affected by the stacked value for any other pixel so the order of processing is irrelevant. These reasons make large parts of the blind stacking pipeline highly parallelisable and thus conducive to GPU processing (pseudo-code algorithms \ref{alg:blind_stacking} and \ref{alg:integration} highlight which computing steps are carried out on the GPU). Using a GPU where possible results in a computational speed up of multiple orders of magnitude. The limitations imposed by the use of a GPU is that data must be first transferred from the CPU to the GPU for processing and then back again for analysis. However, the transfer time is found to be only a very small fraction of the total runtime, thus using a GPU still provides significant speed-up.

There are still potential avenues for reducing this run-time. Firstly, observing at a higher cadence would reduce the length of target streaks and thus, the number of paths that must be searched between frames. This would reduce the runtime required but is not currently possible with our observational set-up. Limiting ourselves to a slower subset of LEO targets would also reduce the number of paths that must be tested and therefore the runtime of the algorithm. This could be done by focusing on targets observed closer to the horizon, or targets on higher orbits (although these approaches bring their own complexities, as mentioned in section \ref{sec:Target velocities}). Additionally, limiting the orbits searched, i.e. focusing on polar orbits only, would drastically reduce the range of directions an object could be moving, limiting the paths searched and thus the runtime. Finally, utilising a higher-performance GPU could result in a computational speed up.

\section{Discussion and conclusions}
\label{sec:Discussion and conclusions}

We have analysed the feasibility of the blind stacking method for the detection of LEO targets. We point out that this study employs the use of real observational data into which simulated signals are injected, thus is more representative than a pure simulation-based approach. The reasons for still employing simulated injections as opposed to a full real-data search campaign are twofold. Firstly, this work is a proof of concept; we wish to prove the feasibility and limitations of the method which is most easily accomplished with a simulated input catalogue. Secondly, detecting real-world targets would require cross-checking against known targets and confirming unknown ones, necessitating additional analysis codes. We plan to follow this work up with a full on-sky implementation of the method.

Current results show that, in terms of recoverability of targets, the method is absolutely feasible, being capable of successfully recovering targets well below the magnitude thresholds available from individual frames. Depending on the exact combinations of optimisation parameters, we have shown this method capable of the recovery of targets at $14^{\rm th}$ magnitude and below. Taking a CubeSat at this sort of altitude to present around $13^{\rm th}$ magnitude shows this method capable of detecting objects with surface areas significantly smaller than this (a change of one magnitude corresponds to a change in surface area, and thus flux, of 2.5 times, under certain assumptions).

The optimal combination of high-level algorithm parameters for the recovery of the faintest objects is currently an exposure time of 0.125s, a pixel binning factor of 4 or 8, and a data-set comprising of $\sim$40 frames. As shown above, this combination leads to the best faint object recovery giving an approximate recovery of 80\% of 13-13.5 magnitude targets and $\sim50$\% of 13.5-14 mag targets. Recoverability of brighter targets is generally consistent at close to 100\% as well.

In terms of run-time we show that the optimal combination of parameters are short exposures, large binning factors, and large numbers of frames per data-set (although this last aspect has a rather limited effect). In order to approach a real-time algorithm, we require exposure times of 0.125s, combined with a binning factor of 4 or 8, or an exposure time of 0.25s, combined with a binning factor of 8. Other combinations, while still sometimes capable of good recovery, generally require too much computational time to be feasible. The best relative run-times are $\lesssim5$, meaning the blind stacking algorithm requires up to 5 times longer than the data collection time to run.

The results presented here have been a proof of concept of the blind stacking procedure, applied to LEO targets. We hope to apply these findings to a full on-sky application in the short term. 




\section*{Acknowledgments}

This work has made use of data obtained using the Warwick CLASP test telescope operated on the island of La Palma by the University of Warwick in the Spanish Observatory del Roque de los Muchachos of the Instituto de Astrofisica de Canarias. BFC acknowledges support from the Science and Technology Facilities Council. JAB acknowledges support from the Defence Science and Technology Laboratory (UK).

\bibliographystyle{jasr-model5-names}
\biboptions{authoryear}
\bibliography{SSA_bib}

\onecolumn

\appendix


\section{Blind stacking algorithm}
\label{app:blind_stacking_pseudocode}

\vspace*{-5mm}

\begin{algorithm}[h!]
\caption{Blind stacking algorithm (Blue lines take place on the GPU)}\label{alg:blind_stacking}
\begin{algorithmic}[1]

\For {Each Pixel}
    \State $\rm PixelLocation \gets [PixelLocationX,PixelLocationY]$
    \State $\rm MasterFrame[PixelLocation] \gets 0$
    \For {Each allowed X Motion}
        \State $x \gets \rm X\ Motion$
        \For {Each allowed Y Motion}
            \State $y \gets \rm Y\ Motion$
            \State ${\rm Total\ Motion} \gets \sqrt{x^2 + y^2}$
            \If {Total motion within allowed range}
                \color{blue}
                \State $\rm PixelValue \gets 0$
                \For {Each Frame}
                    \State $\rm PixelLocationX \gets PixelLocationX + x$
                    \State $\rm PixelLocationY \gets PixelLocationY + y$
                    \State $\rm PixelLocationFrame \gets [PixelLocationX,PixelLocationY]$
                    \State $\rm PixelValue \gets PixelValue + Frame[PixelLocationFrame]$
                \EndFor
                \If {$\rm PixelValue>MasterFrame[PixelLocation]$}
                    \State $\rm MasterFrame[PixelLocation] \gets PixelValue$
                \EndIf
                \color{black}
            \EndIf
        \EndFor
    \EndFor
\EndFor

\end{algorithmic}
\end{algorithm}

\vspace*{-5mm}


\section{Integration algorithm}
\label{app:integration_pseudocode}

\vspace*{-5mm}

\begin{algorithm}[h!]
\caption{Integration algorithm (Blue lines take place on the GPU)}\label{alg:integration}
\begin{algorithmic}[1]

\For {Each Pixel}
    \State $\rm PixelLocation \gets [PixelLocationX,PixelLocationY]$
    \State $\rm IntegratedFrame[PixelLocation] \gets 0$
    \For {Each allowed X Motion}
        \State $x \gets \rm X\ Motion$
        \For {Each allowed Y Motion}
            \State $y \gets \rm Y\ Motion$
            \State ${\rm Total\ Motion} \gets \sqrt{x^2 + y^2}$
            \If {Total Motion within allowed range}
                \color{blue}
                \State $\rm NSteps \gets Total\ Motion$
                \State $\rm PixelValue \gets 0$
                \For {Each Step}
                    \State $\rm PixelLocationX \gets PixelLocationX + x/NSteps$
                    \State $\rm PixelLocationY \gets PixelLocationY + y/NSteps$
                    \State $\rm StreakPixel \gets [PixelLocationX,PixelLocationY]$
                    \State $\rm PixelValue \gets PixelValue + Image[StreakPixel]$
                    \State $i \gets i+1$
                \EndFor
                \If {$\rm PixelValue>IntegratedFrame[PixelLocation]$}
                    \State $\rm IntegratedFrame[PixelLocation] \gets PixelValue$
                \EndIf
                \color{black}
            \EndIf
        \EndFor
    \EndFor
\EndFor

\end{algorithmic}
\end{algorithm}

\end{document}